\RequirePackage[l2tabu, orthodox]{nag}
\documentclass[%
twocolumn,citeautoscript,
nofootinbib,
amsmath,amssymb,
floatfix,
]{revtex4-1}
\usepackage{times}
\usepackage{graphicx}
\usepackage{dcolumn}
\usepackage{bm}
\usepackage{hyperref}
\usepackage{etoolbox} 
\usepackage{tabularx,tabulary}
\usepackage[capitalize]{cleveref}
\usepackage{xcolor}
\hypersetup{
	colorlinks,
	linkcolor={blue!100!black},
	citecolor={blue!100!black},
	urlcolor={blue!100!black}
}
\makeatletter

\appto{\appendix}{%
	\@ifstar{\def\theequation@prefix{A.}}%
	{}%
}
\makeatother
\usepackage{color,float}
\usepackage{chngcntr}
\newcommand{\hlt}[1] {\textcolor{black}{#1}}
\newcommand{\sk}[1] {\textcolor{black}{#1}}
\newcommand{\etal}{\textit{et al}.}
\usepackage[pagewise,mathlines]{lineno}

\begin{document}
	
	
	\title{Shock related unsteadiness of axisymmetric spiked bodies in the supersonic flow}
	
	\author{Devabrata Sahoo}
	\email{Contributed equally to this work as first authors}
	\author{S. K. Karthick$^*$}%
	\altaffiliation[Corresponding author: ]{skkarthick@ymail.com}
	\author{Jacob Cohen}%
	\affiliation{%
		Faculty of Aerospace Engineering, Technion-Israel Institute of Technology, Haifa, 3200003, Israel
	}%
	
	\author{Sudip Das}
	\affiliation{
		Department of Space Engineering \& Rocketry, Birla Institute of Technology, Mesra, 835215, Jharkand, India 
	}%
	
	\date{\today}
	
	\begin{abstract}
		Shock related unsteadiness over axisymmetric spiked body configurations is experimentally investigated at a freestream supersonic Mach number of 2.0 at 0$^\circ$ angle of attack. Three different forebody configurations mounted with a sharp spike-tip ranging from blunt to streamlined (flat-face, hemispherical, and elliptical) are considered. Steady and unsteady pressure measurements, short-exposure high-speed shadowgraphy, shock footprint analysis from $x-t$ plots, and identification of dominant spatiotemporal modes through modal analysis are carried out to explain the unsteady flow physics. The present investigation tools are validated against the well-known events of `pulsation' associated with the flat-face case. The hemispherical case is characterized by the formation of a separated free shear layer and associated localized shock oscillations. The cycle of charging and ejection of fluid mass from the recirculation zone, confined between the separated shear layer and the spiked body, is identified to drive the flow unsteadiness. Such an event triggers the out-of-phase motion between the separated and reattachment shocks. In the elliptical case, the overall flow field resembles that of the hemispherical case, except with dampened unsteadiness. The value of the cone angle ($\lambda$) associated with the recirculation region is found to be responsible for the \sk{fluctuations} from the charging and ejection of fluid mass. Thereby it controls the extent of out-of-shock phase motion. In the elliptical case, $\lambda$ is observed to be smaller and exhibits a reduction in shock unsteadiness. \hlt{Based on the gathered results and understanding, the reduction in unsteadiness associated with the aerodisk mounted on the hemispherical forebody is explained via the almost complete elimination of the out-of-phase shock motion.}
	\end{abstract}
	
	\pacs{Valid PACS appear here}
	\maketitle
	
	
	\section{Introduction} \label{intro}
	
	Flow unsteadiness is generally observed ahead of a variety of axisymmetric forebodies, flying especially at supersonic and hypersonic speeds \citep{247182:10473542, 247182:6199877}. Configurations like mixed compression inlets \citep{247182:10473724,247182:10474885}, double cones \citep{247182:10473930,247182:10453234}, forward-facing cavities \unskip~\citep{247182:10473931}, axially positioned cavities \unskip~\citep{247182:10453233}, wall protrusions \citep{247182:5719032}, and spiked/aerodisk forebodies \citep{247182:5562223,247182:5562224,247182:5562232} are known to exhibit severe flow unsteadiness. The unsteadiness is seen throughout a wide range of flow Reynolds numbers ($Re$) corresponding to laminar as well as turbulent states \citep{247182:5562230,247182:5562233,247182:5562234}. In general, the strongest form of shock-related unsteadiness is termed as `buzzing' \citep{247182:10474137,247182:10474228}. It is primarily driven by the inviscid unsteady shock phenomena \citep{247182:10452954} and is known to be devastating to the vehicle structure. Another form of unsteadiness is mostly dominated by the viscous shockwave boundary layer interactions (SWBLI), which are found to be less intense in comparison with the former \citep{247182:10474441,247182:10474525}. However, these interactions could not be neglected in practice, as their occurrence is still a threat to the vehicle's structural integrity and flight control. 
	
\begin{figure}
		\centering 
		{\includegraphics[width=0.8\columnwidth]{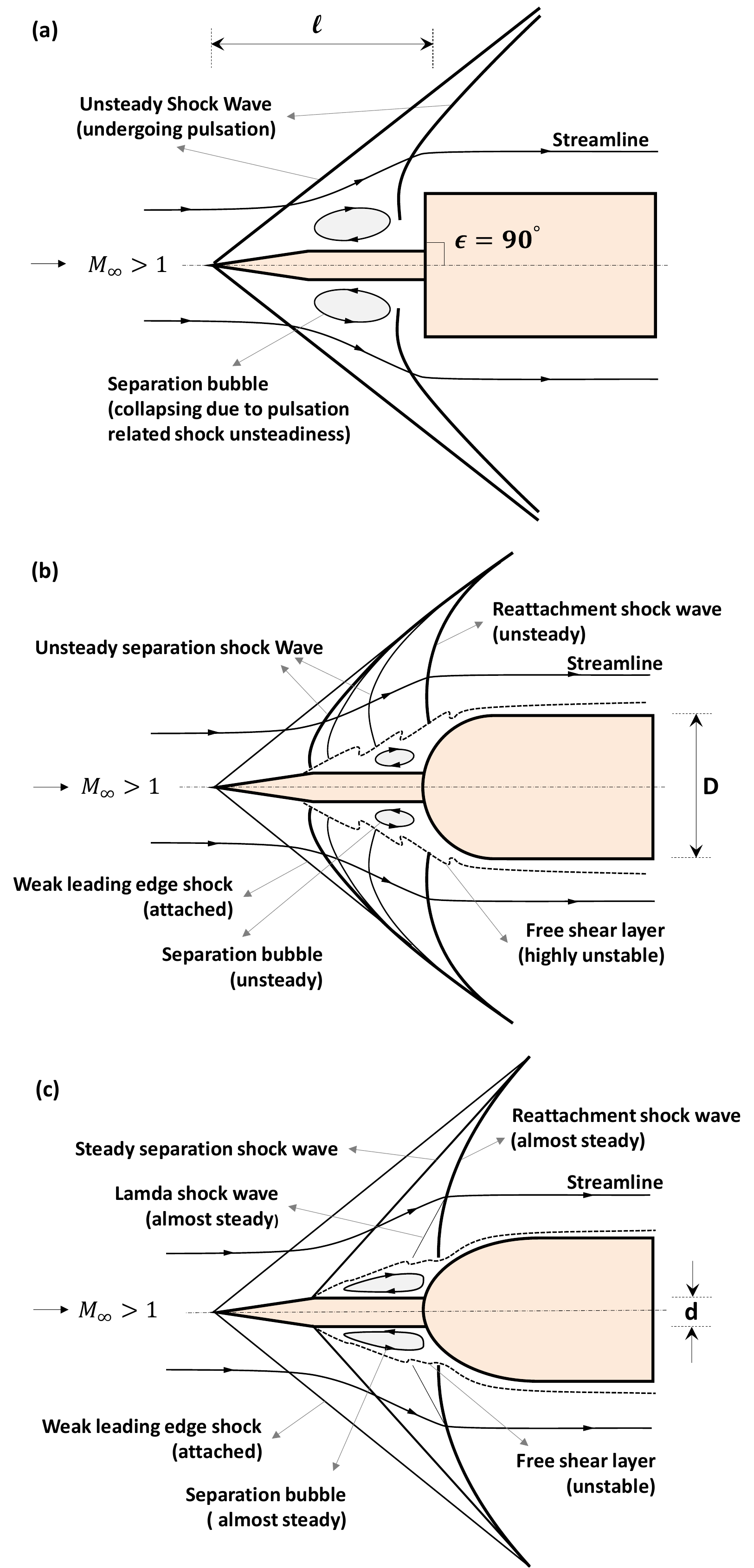}}
		\caption{\hlt{Schematic showing the basic flow features encountered at a given instant over a spiked forebody of different forebody configurations (a) spiked flat-face forebody (blunt), (b) spiked hemispherical forebody (less blunt/streamlined), and (c) spiked elliptical forebody (streamlined) for $M_\infty>1$. Flow is from left to right.}}
		\label{fig:1}
	\end{figure}

	In the middle of the last century \citep{247182:5562220,247182:10452955,247182:5562223}, significant studies had been conducted to design space launch vehicles, which can fly at higher speeds with minimal drag coefficient ($c_d$). Spiked axisymmetric forebodies of specific forebody shape of base diameter ($D$) were found to be more efficient and practical \citep{247182:10453038}, among the several proposed configurations. The forebody shapes were ranging from a simple flat-face cylinder to a more sophisticated ogives or rounded-off cylinders \citep{247182:10452955}, as shown in Figure~\ref{fig:1}. With the attachment of a spike, those forebody shapes had achieved a reduced drag of up to 68\% \citep{247182:5562222}. Spiked bodies exhibit the strongest form of flow unsteadiness termed `pulsation' \citep{247182:5562226,247182:5562227}, which under particular flow and geometrical conditions is similar to the `buzzing' phenomena as observed in a typical supersonic inlet \citep{247182:10474228}. \sk{The intensity of flow unsteadiness (in general defined by calculating the root mean square of the fluctuations in pressure, velocity, and changes in shocks' position) is reported to be remarkedly reduced for the streamlined spiked forebodies \citep{247182:5562232,247182:5562234} and for the forebodies with aerodisk type spikes \citep{247182:5562221,247182:10453264}.} 
	
	The underlying flow physics that govern the varying intensity of unsteady shock-laden flow field is still a topic of interest, even though flying configurations were realized through several experimental \citep{247182:5562230} and computational \citep{paskonov} studies. Many researchers have found (see Table \ref{tab:1}) that there exist a variety of parameters influencing the unsteady flow field around the spiked bodies including Reynolds number ($Re_D$), freestream Mach number ($M_\infty$), spike length to forebody diameter ratio ($l/D$), spike diameter to forebody diameter ratio ($d/D$), and the shape of the spike and the forebody themself. Many such parametric studies had been listed from both experiments and computations, in the review paper of \cite{247182:5562221}.

	Since its discovery, researchers \citep{247182:5562223,247182:10081462,247182:5562225,247182:5562224} had been pondering to find a way to cope with the extent of observed unsteadiness, especially in the spiked flat-face body as it is known to exhibit a very violent form of shock-related unsteadiness. The associated strongest forms of unsteadiness were then identified as `pulsation' ($0.5<l/D<1.5$) and `oscillation' ($1.5<l/D<2.5$) \citep{247182:5876290,247182:10453043,247182:5562226,247182:5562227,feszty}, based on the spike length ($l$) and the model base-body diameter ($D$) for $M_\infty>1$. In the oscillation mode, the shock moves laterally, while in the pulsation mode, it moves axially.
	
	As the first experimental observations of unsteady events happened only in the laminar flow regime at high Mach number \citep{247182:5562224}, seeking a numerical approach was considered to be an ideal way to understand the unsteady flow physics. Unsteady laminar compressible Navier-Stokes equations were solved numerically to resolve the flow field around the spiked bodies both spatially and temporally \citep{247182:10453378,247182:10453046}. At a lower Reynolds number ($Re_D\approx0.1\times 10^6$), for certain geometrical configurations, `pulsation' and `oscillation' phenomena were seen independently. Switching between the two forms in the same run while varying the spike length was reported in \citep{247182:10453042,247182:10474608}. Recently, the unsteady flow field around the spiked bodies has also been studied for higher Reynolds numbers using appropriate turbulence modeling \citep{247182:10453150,247182:10453124}. The reported results are in agreement with the proposed mechanism of `pulsation' and `oscillation' provided by \cite{247182:10474718,247182:10474608}, and \cite{247182:5562227,feszty}. However, in the experiments at higher Reynolds numbers ($Re_D\approx1-7\times 10^6$), researchers \citep{247182:10081462} reported the absence of `oscillation' form of unsteadiness at different $d/D$ and $l/D$.
	
	\sk{Many investigators have reported the usage of novel active and passive control techniques on the spiked bodies to achieve stable or reduced unsteady flow with superior heat transfer and drag reduction capabilities. Many such studies are listed in the review article of \cite{247182:10453047}, elaborately. Owing to page limitations and to keep the discussions within the scope of the present paper, only a few of the relevant papers pertaining to the forebody shape changes (passive control) to reduce shock-related unsteadiness are briefly reviewed.}
	
	\begin{table*}
		\centering
		\caption{\hlt{Some of the relevant list of experiments and numerical simulations performed in the past and present studies at different flow conditions, model configurations, and investigation techniques.}}
		\label{tab:1}
		\rotatebox{0}{				
			\resizebox{\textwidth}{!}{
				\begin{tabular}{@{}cccccccccc@{}}
					\toprule \textbf{References} & \textbf{Forebody} & \textbf{Spike} & \textbf{$\boldsymbol{[l/D]}$} & {$\boldsymbol{[d/D]}$} & \textbf{$\boldsymbol{M_\infty}$} & \textbf{$\boldsymbol{Re_D \times 10^6}$} & \textbf{$\boldsymbol D$} $(mm)$ & \textbf{Methods} & \textbf{Tools} \\
					Mair \cite{247182:5562223} & F, H & ST & 0 – 2.1 & 0.1 & 1.96 & 0.165 & 12.7 & E & S, P \\
					Beastall \& Turner \cite{247182:10081462} & F & ST, CA & 0 – 4.6 & 0.16 & 1.5, 1.6, 1.8 & 1.125 & 38.1 & E & S, P \\
					Bogdonoff \& Vas \cite{247182:5562225} & F, H & ST & 0 – 8.0 & 0.1 & 14 & 0.365 & 12.7 & E & S, P \\
					Maull \cite{247182:5562224} & F, H & ST & 0 – 4 & 0.06 & 6.8 & 0.085 & 12.7 & E & S \\
					Antonov \etal \cite{antonov} & F & ST & 0 – 5 & 0.06 & 2.1 – 6 & 0.07 – 1.6 & - & E & S \\
					Guenther \& Reding \cite{247182:5562230} & HC & FA & 0.86 & 0.062 & 1.6 – 3.5 & 4.825 – 7.046 & 795.1 & E & S, P \\
					Kenworthy \cite{247182:5562226} & F, C & ST & 0 – 2.5 & 0.06 & 2.21, 6 & 0.12, 0.13 & 46, 20 & E & S, P \\
					Shang \etal \cite{shang} & CF & BT & 0.14 – 0.62 & 0.14 & 3 & 0.62 & 88.65 & N & UL \\
					Calarese \& Hankey \cite{247182:10453042} & CF & BT & 0.14 – 0.62 & 0.14 & 3 & 0.62 & 88.65 & E & S, P \\
					Mikhail \cite{247182:10453046} & F & FT & 0.68 – 1.06 & 0.177 & 1.72 & 2.322 & 64.77 & N & UR \\
					Khlebnikov \cite{khlebnikov} & F, C & ST, FA & 0.35 – 2.02 & 0.06 & 3 & 0.8 & 60 & E & T, P \\
					Yamauchi \etal \cite{yamauchi} & H & ST & 0.5,1.0,2.0 & 0.1 & 2.01,4.15,6.8 & 0.14 – 1.4 & 76.2 & N & SL \\
					Mehta \cite{mehta} & H & ST & 0.5, 1.0, 2.0 & 0.1 & 6.8 & 0.14 & 76.2 & N & UR \\
					Feszty \etal \cite{feszty} & F & ST & 2 & 0.065 & 6 & 0.13 & 20 & N & UL \\
					Feszty \etal \cite{247182:5562227} & F & ST & 1 & 0.065 & 2.21 & 0.12 & 46 & N & UL \\
					Jagadeesh \etal \cite{jagadeesh} & BC & ST, IA & 1 & 0.052 & 6.99 & 0.123 & 50 & E, N & S, UL \\
					Zapryagaeva \& Kavun \cite{zapryagaeva} & F & ST & 0.55 – 1.45 & 0.16 & 6.01 & 0.65 & 50 & E & S, P \\
					Zapryagaev \& Kavun \cite{zapryagaevb} & F & ST & 1 & 0.16 & 6.01 & 0.65 & 50 & E & S, P \\
					Panaras \& Drikakis \cite{247182:5562232} & F, H, CF, C & ST, BT & 0.5 – 2.5 & 0.12 & 3.0 – 6.8 & 0.12 – 0.394 & 20 – 88.65 & N & UL \\
					Ahmed \& Qin \cite{247182:5562221} & H & ST, HA & 0.5 – 2.5 & 0 – 0.4 & 6 & 0.037 & - & N & UL \\
					Sahoo \etal \cite{247182:5562234} & H & ST, BT, IA & 0.75, 1, 1.5 & 0.133 & 2.0 & 0.35 & 15 & E, N & S, P, UL \\
					Balakalyani \& Jagadeesh \cite{247182:10453250b} & F & ST & 0.7, 1.8 & 0.07 & 6 & 0.414 & 70 & E & S, P \\
					Xue \etal \cite{247182:10453124} & H & FA & 1 & 0.07 & 4.5 & 0.5 & 70 & N & DES \\
					Konstantin \cite{247182:10453150} & F & ST, FT & 0.5 – 1.5 & 0.065 & 2.22 & 1 & 23 & N & UR \\
					Sahoo \etal \cite{247182:10453264} & F, H & ST, HA & 1 & 0.06 & 2.0 & 2.159 & 50 & E, N & S, P, UL \\
					\textbf{Present Studies} & \textbf{F, H, EL} & \textbf{ST} & \textbf{1} &\textbf{ 0.06} & \textbf{2.0} & \textbf{2.159} & \textbf{50} & \textbf{E} & \textbf{S, P, M} \\
					\multicolumn{10}{l}{All the experiments or simulations were performed at zero degree angle of attack. In addition, the considered models are axisymmetric}\\ 
					\multicolumn{10}{l}{in nature and exhibited unsteady flow features; F – flat-faced; H – hemispherical; C – conical; CF – cone frustum; HC – hemispherical cone;}\\
					\multicolumn{10}{l}{BC – blunt cone; EL – elliptical; ST – sharp tip; BT – blunt tip; FT – flat tip; CA – conical aerodisk; FA – flat aerodisk; IA – inverted}\\
					\multicolumn{10}{l}{conical aerodisk; HA – hemispherical aerodisk; E – experiments; N – Numerical simulations; S – shadowgraph or schlieren; P – steady or }\\
					\multicolumn{10}{l}{unsteady pressure; U(S)L – unsteady(steady) laminar Navier-Stokes simulations; UR – unsteady Reynolds-averaged Navier-Stokes}\\
					\multicolumn{10}{l}{simulations; DES – Detached Eddy Simulations; $n.a$ – not available; T – surface tufts; M – Modal analysis}\\ \toprule
		\end{tabular}}}
	\end{table*} 
	
	\sk{In the pioneering work of Moeckel\unskip~\cite{247182:5939825}, the author systematically reported a range of flow unsteadiness for wall-mounted protrusions having a sharp corner in a supersonic flow. The flow field around the two dimensional model resembles the axisymmetric flat-face spiked bodies. The author reported the existence of unsteady flow field for particular geometrical configurations similar to spiked bodies as the protrusion height was gradually increased. However, his study was just confined to protrusion heights in the order of the boundary layer thickness. Unlike in Moeckel's case, in spiked bodies, separation happens in the leading edge or close to it, where the boundary layer is comparatively thin.}
	
	\sk{In the subsequent studies by Maull \cite{247182:5562224} on the axisymmetric spiked bodies, the influence of geometrical parameters and the forebody shapes were considered in order to identify the stable flow zones. From his experiments, the author concluded that the streamlined bodies like ellipsoids of revolution with shorter spike tend to achieve stable flow with reduced drag coefficient.}
	
	\sk{Antonev \etal \cite{antonov} had also stated that the unsteadiness in spiked bodies were observed to decrease substantially by changing the geometrical features of the spiked body itself. Based on schlieren and shadowgraph images captured at higher frame rates (fps) such kind of unsteady flow control were qualitatively demonstrated. Later, three unsteady pressure transducers were mounted on the spiked bodies at a fixed radial location and different azhimuthal positions to measure the varying intensity of flow unsteadiness \citep{247182:10453042}. The unsteady pressure data revealed the dominant temporal details, despite providing minimal information on the spatial modes due to poor resolution in sensor placements.} 
	
	\sk{In the work of Kenworthy \citep{247182:5562226}, the importance of the forebody cone angle was discussed. Blunt cone with a spike produced a stronger unsteady flow field whereas shallow cone angle forebody with a spike was observed to be almost stable. In the studies of Panaras \& Drikakis \citep{247182:5562232}, double-cone forebody shapes were varied and the dominant unsteady flow modes were observed to be switching based on the first cone angle. Here, the first cone resembles a spike of varying geometry inserted in front of a conical forebody.}
	
	In addition, in a few of the cases \citep{247182:5562223,247182:5562224,247182:10453264}, the pulsation form of unsteadiness vanishes abruptly as the spiked forebody shape changes to a hemisphere. The importance of longer spikes and rounded-off or streamlined forebodies to achieve better drag reduction were reported by \cite{247182:10452955} using time-averaged measurements. Later, researchers had attempted to study the unsteady flow field around the rounded-off edges. In the work of \cite{247182:10453233}, the authors had observed an abrupt switch in the unsteadiness from unstable to stable flow field as the flat-face model was rounded-off or streamlined progressively along the corners.
	
	\hlt{In the recent paper of \cite{247182:10453250}, the influence of the spike-shape and shoulder-rounding on the modes of unsteadiness have been studied in a blunt spiked body through unsteady pressure measurements in a hypersonic shock-tunnel. More detailed studies on the spike-shape and flow unsteadiness was reported by some of the authors in \cite{247182:5562234}, both experimentally and numerically. Similarly, in a recent work of some of the authors in \cite{247182:10453264}, supersonic wind tunnel experiments of flat-face and hemispherical forebody shapes with a spike reveal a significant drop in the intensity of the shock-related unsteadiness.}
	
	\sk{The influence of the forebody shapes is not surprising, as the shapes determine the net drag force acting, heat transfer capabilities, and payload bearing capacities. However, for most of the space missions, the payload's storage in the forebody is given priority, which always requires a higher volume. Reducing the volume in the forebody region leads to the allocation of an extra space somewhere else in the vehicle, thereby increasing the net weight of the vehicle itself. For example, the overall forebody volume in comparison with a simple flat-face cylindrical forebody is reduced by 22\% and 33\% for hemispherical and elliptical forbodies, respectively.}
	
	\sk{In summary, the following aspects are evident from the brief literature review of the spiked bodies: a. The existence of a specific form of unsteadiness is widely known for a given spiked body configuration and flow conditions; b. Unsteadiness is accessed through quantitative temporal measurements; c. Unsteadiness could be controlled using passive control techniques like modifying the geometrical parameters of the spiked bodies; d. Shorter spike and streamlined forebody achieve almost a stable flow; e. Streamlined spiked forebody poses payload placements issues and aero-thermodynamic design challenges.}
	
	\sk{In this context, we aim to study the unsteady flow field as we change the spiked forebody shapes from a blunt forebody like a flat-face cylinder to a streamlined forebody like an ellipsoid. In particular, the main objectives are a. To extract the dominant spatiotemporal flow events; b. To identify the critical flow features contributing to the varying flow unsteadiness; c. Utilize the gathered knowledge to frame generic control guidelines in mitigating the shock-related flow unsteadiness observed in similar bodies for efficient flight design.} 
	
	\sk{It should be noted that the coupled spatiotemporal analysis based on the conventional shadowgraph imaging and the modern image decomposition tools (Proper Orthogonal and Dynamic Mode Decomposition tools) presented in this paper is a unique approach utilized to understand the unsteady flow physics in the spiked body flows. To the best of our knowledge, such an approach applied to supersonic spiked body flows, has not been attempted before.}
	
	\begin{figure*}
		\centering{\includegraphics[width=0.7\textwidth]{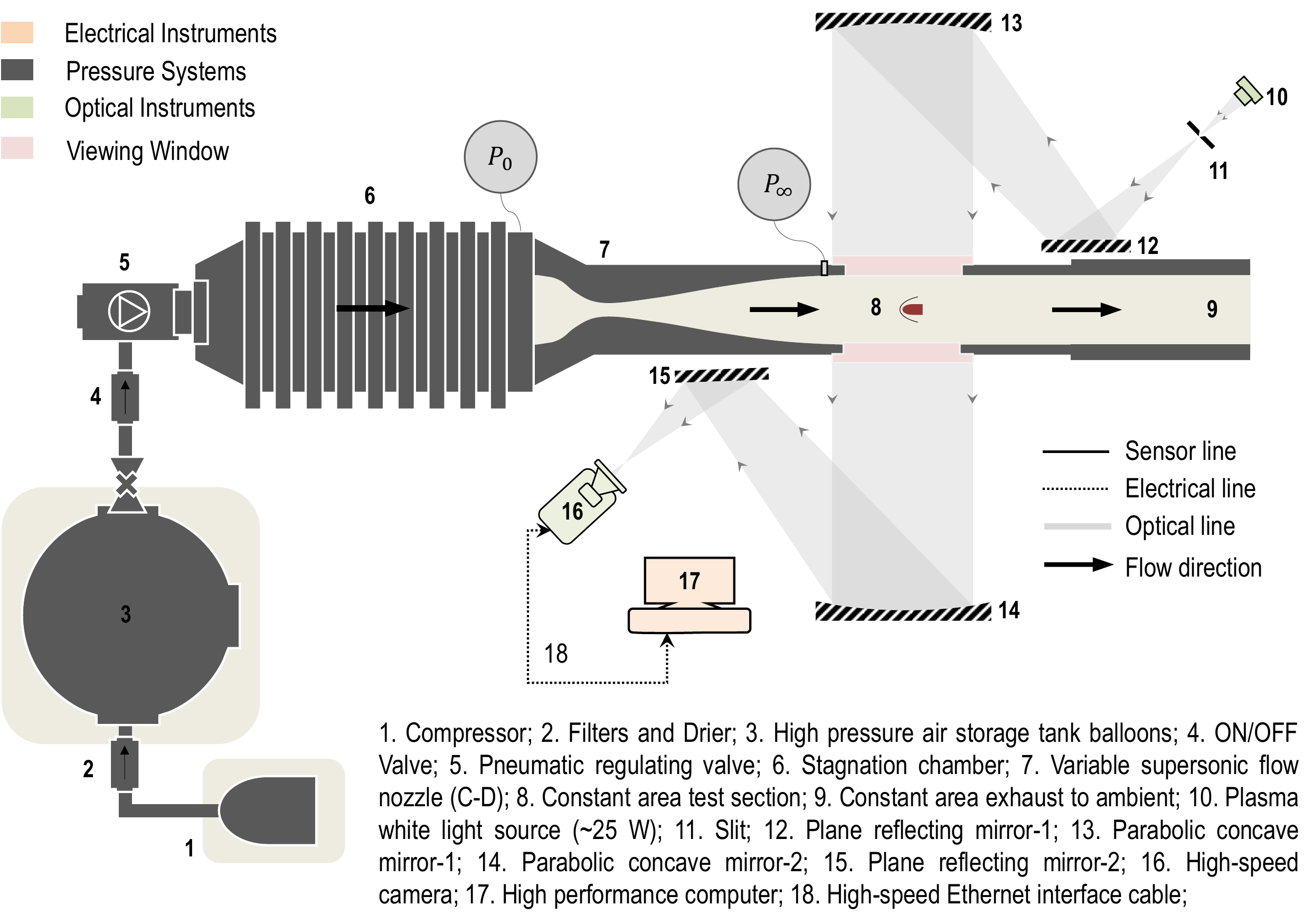}}
		\caption{\hlt{Schematic of the supersonic blow-down type wind tunnel with a `Z-type' high-speed short-exposure shadowgraph imaging \cite{247182:5609367} setup to study the unsteadiness in different spiked-body configurations at $M_\infty=2.0$. Flow is from left to right.}}
		\label{fig:4}
	\end{figure*}
	
	The flow unsteadiness on a variety of geometrical configurations and flow conditions are listed in Table~\ref{tab:1}, based on the comprehensive set of literature. Most of the research works are numerical studies on simplistic configurations at a relatively lower Reynolds number and a higher Mach number ($M_\infty>4 $). \hlt{Hence, in contrast to the literature and in continuation with our previous studies, we have selected a generic case of a sharp spike-tip flat-face (ST-F) and a sharp-spiked hemispherical forebody (ST-H), primarily for experimental studies at a higher Reynolds number ($Re_D\sim2\times 10^6$) and a lower Mach number ($M_\infty\sim2 $).} 
	
	\hlt{At first, shock-related unsteadiness associated with the sharp-tip spike mounted flat-face forebody configuration (ST-F) is studied using various experimental and data analysis techniques. The methodologies are then validated with the well-known event of `pulsation' observed in ST-F. Later, hemispherical (ST-H) and elliptical (ST-E) forebodies mounted with a sharp-tip spike are investigated to study the extent of forebody rounding effects on the shock-related unsteadiness. In addition, with the gathered knowledge, a simple hemispherical spike/aerodisk tip on a hemispherical forebody configuration (HA-H) \cite{247182:5562221,247182:10453264} is investigated, as it is known to exhibit lower flow unsteadiness and drag. The flow physics behind the aforementioned behavior is then explained. Finally, generic control methods for shock-related unsteadiness based on the outcome of the present study are proposed.}
	
	\hlt{The manuscript is organized as follows. A detail description of the experimentation procedure is given in $\S$\ref{expt} following the introduction. It includes details about the facility ($\S$\ref{facility}), geometrical aspects of the selected configurations ($\S$\ref{geometry}), and a brief layout of the measurement methodology ($\S$\ref{meas.method}). Uncertainties involved in the current experimental procedures and data analysis are given in $\S$\ref{uncert.}. $\S$\ref{res.disc.} contains the results and discussions, where the drag reduction of the adopted test models are presented first ($\S$\ref{drag}). The observed shock-related unsteadiness on the considered models is then addressed using the unsteady pressure measurement data ($\S$\ref{spectra}). Then, the dominant energetic and dynamic spatiotemporal contents present in these configurations are described, and the unsteady shock-laden flow field is explained ($\S$\ref{rd3}). Based on the gathered understandings, an aerodisk spike configuration is investigated subsequently to explain the reduced shock-related unsteadiness and to provide control guidelines in $\S$\ref{case4} before reaching the conclusions ($\S$\ref{conc.}). In the appendix ($\S$\ref{data_an}), the data analysis procedures to extract the dominant spatiotemporal modes from the modal analysis of shadowgraph images are briefly described and validated. In addition, in the appendix ($\S$\ref{flow_flat-face}), a brief summary on the flow features observed around the well-known flat-face spiked configuration is provided for completeness. In the supplementary, high-speed shadowgraph videos and some numerical results\footnote{Filename of the experimental shadowgraph videos given for different spiked forebody configurations in the supplementary: \href{https://youtu.be/nYG-BfhB02A}{`video1'} (Flat-face), \href{https://youtu.be/QCdHLPZE6k0}{`video2'} (Hemisphere), \href{https://youtu.be/v1UD0pAH_LM}{`video6'} (Elliptical), and \href{https://youtu.be/5Ha7kJerZWg}{`video7'} (Aerodisk). Higher frame rate videos for the hemispherical configuration is available in \href{https://youtu.be/2LAdCpIQAP8}{`video4'} and \href{https://youtu.be/gHpjJcdnB58}{`video5'}. Detached-Eddy Simulation (DES) numerical results showing the video of instantaneous numerical schlieren is given in \href{https://youtu.be/EEVu-MS4-X8}{`video3'}. Brief captions for each of the video is available in the `VideoCaptions.txt' file.} are provided to visually appreciate the varying unsteady flow field around the axisymmetric spiked bodies.}
	
	\section{Experimentation}\label{expt}
	All the experiments are performed in the blow-down supersonic wind tunnel facility located in the Wind Tunnel Complex of Technion-Israel Institute of Technology, Israel (see Figure \ref{fig:4}). The details of the experimental facility, the measurement methodology, and the operating conditions adopted in the present work are provided in the following sub-sections.
	
	\subsection{Supersonic blow-down facility}\label{facility}
	The supersonic wind tunnel having a test section size of 400 mm (width) $\times$ 500 mm (height) is designed with a variable throat nozzle (convergent-divergent) to generate a freestream flow of Mach number ($M_\infty$) ranging from 1.6 to 3.5. The high-pressure air required to run the supersonic tunnel is taken from the compressed air storage. The storage facility consists of 48 balloons having a compressed air storage volume of $0.6$ $m^3$/balloon. These balloons can be charged up to a maximum storage pressure of 200 bar. The storage facility is charged using two units of 5-stage piston-type air compressors, operating in parallel. Each compressor has a rated power of 400 kW and has an air discharge capacity of 0.5 $m^3/s$. Before storing the compressed air, moisture and impurities are removed using appropriate filter systems. The pneumatic control valve and ON/OFF gate valve are used to regulate the compressed air flow entering the settling chamber. Freestream flow stagnation conditions are measured at this location. Then, the compressed air is passed through the convergent-divergent (C-D) nozzle corresponding to a particular $M_\infty$. At the end of the C-D nozzle, freestream flow conditions are established in the constant area test section of length 1 m.  Optical quality glass windows (BK-7) (550 mm in length and 300 mm in height) are mounted for flow visualization studies on the sidewalls of the test section. Suitable sting arrangements are made inside the test section to mount the models and the force and pressure measurement systems. The air is then discharged into the ambient through a constant area duct of similar size as that of the test section. A typical schematic explaining the layout of the supersonic blow-down type wind tunnel facility is shown in Figure \ref{fig:4}.
	
	\begin{figure*}
		\centering{\includegraphics[width=0.9\textwidth]{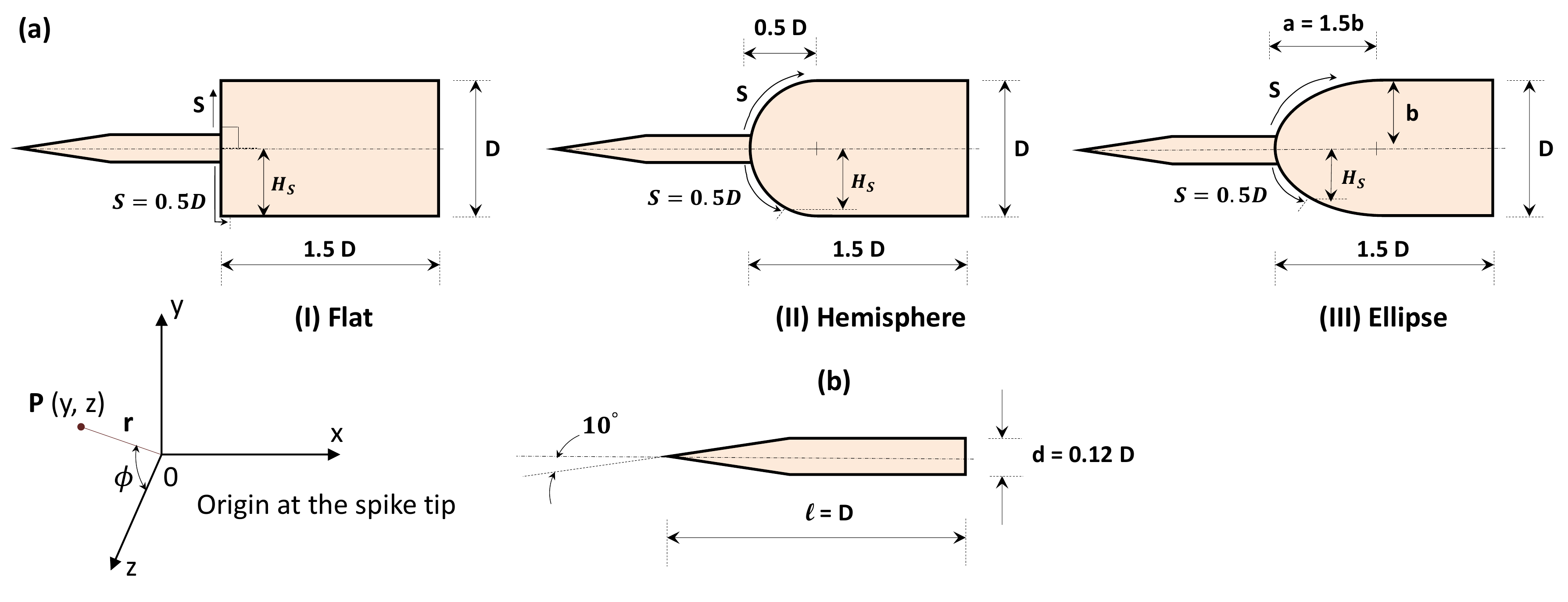}}
		\caption{\hlt{Schematic showing the primary geometrical parameters required to construct different spiked body configurations: (a) spiked forebody shapes, and (b) spike-tip shape. Origin is always at the tip of the spike and faces against the flow. Azimuthal location of an arbitrary point anywhere on the model surface is represented using the circumferential angle ($\phi$). Distance of any point along the model surface in the axial direction is given by $S$.}}
		\label{fig:5}
	\end{figure*}
	
	\subsection{Model geometry}\label{geometry}
	A cylindrical model with a base diameter ($D$) of 50 mm and an overall length of 1.5D has been adopted for all the configurations investigated in the present work. For the elliptical forebody configuration, a semi-major axis length of $a=0.75D$ and a semi-minor axis length of $b=0.5D$ has been employed. A sharp spike having a semi-cone angle of $10^{\circ}$ and a stem diameter of $d=6$ mm ($0.12D$) has been used. The spike length ($l$) is chosen to be equal to that of the base body diameter ($D$). The cross-sectional area of the test model results in only 0.98\% blockage in the test section, provided the angle of attack is kept at $\alpha=0^{\circ}$. Figure~\ref{fig:5} shows the primary geometrical details of the spiked body configurations. Any points on the model surface at a given $xy-$plane is measured in terms of a segment length parameter ($S$) along the model surface from the stagnation point of the blunt body without the spike. Any points on the model surface at a given $yz-$plane are represented by an azimuthal angle $\phi$ as shown in Figure~\ref{fig:5}b. A parameter called the geometrical shape factor ($\xi$) is proposed to distinguish between the blunt and streamlined forebodies. The parameter ($\xi $) is defined as the ratio of vertical distance ($H_S$) measured from the forebody axis to the point where $S \approx 0.5D$ along the forebody base of radius ($D/2$) as shown in Figure \ref{fig:5}. In case of the flat-face spiked body configuration $\xi_F=2H_S/D\approx1 $. For the hemispherical and elliptical spiked body configurations $\xi_H\approx0.92$ and $\xi_E\approx0.84$, respectively. The definition of $\xi $ comes handy in the discussions made in the subsequent sections.
	
	\subsection{Measurement methodology} \label{meas.method}
	The wave drag associated with the three different models is measured using an in-house strain gauge balance.  \sk{In the context of shock-related unsteadiness, we focus on identifying the frequencies associated with shocks' motion from the $x-t$ analysis of time-resolved shadowgraph images and obtaining the pressure spectra from the Fast Fourier Transform (FFT) of unsteady pressure sensor signals.} At first, shadowgraph is utilized as a qualitative imaging technique. Later, the light intensity fluctuations from the shadowgraph images are used to get information regarding the extent of shock-related unsteadiness through rigorous image processing and data analysis (see $\S$\ref{data_an} given in the appendix). Quantitative information like the shock trajectories from $x-t$ plots, and spatiotemporal modes obtained from the modal analysis are then acquired. The details of all these measurement procedures are given briefly in the following sub-sections.
	
	\begin{figure*}
		\centering{\includegraphics[width=0.9\textwidth]{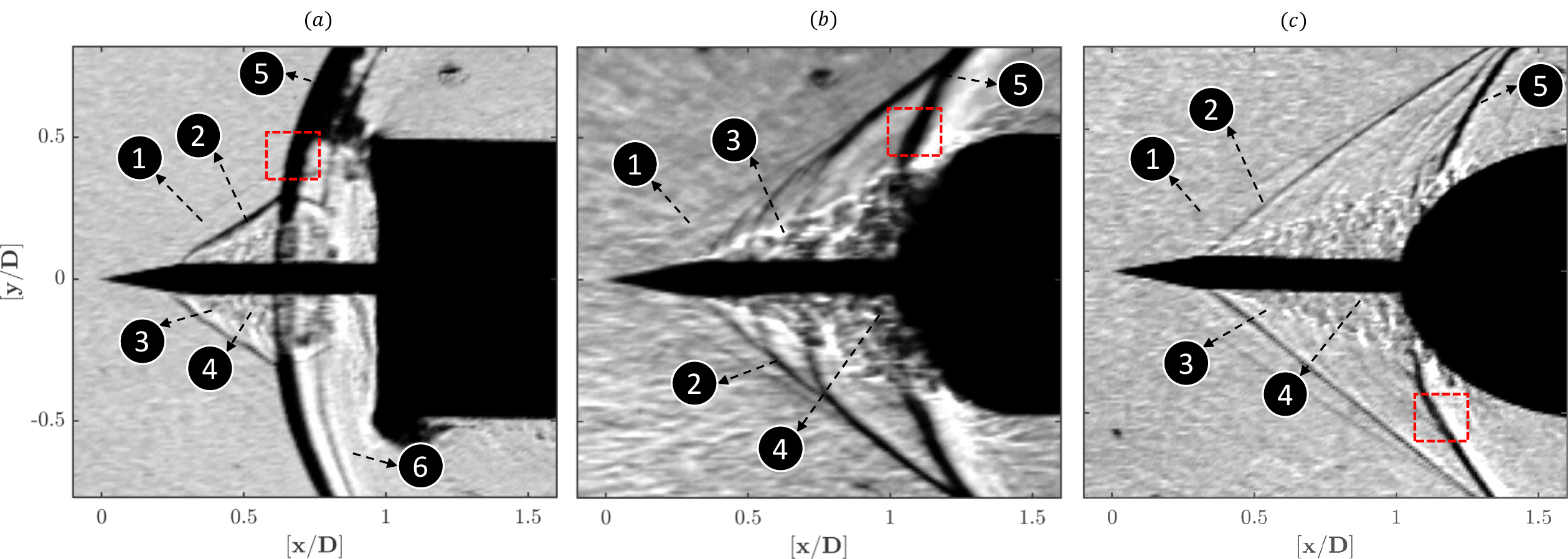}}
		\caption{{\hlt{Typical instantaneous shadowgraph images showing the basic flow features observed around different spiked forebody configurations at $M_\infty=2.0$. Dominant flow features: 1. Weak leading edge shock; 2. Separation shock (SS); 3. Separated shear layer; 4. Recirculation region; 5. Reattachment shock (RS); 6. Refracted shock; The thickness of the shock seen inside the red dotted boxes qualitatively represents the shock strength. Flow is from left to right. High-speed shadowgraph videos for each of the cases are available in the supplementary (flat-face: \href{https://youtu.be/nYG-BfhB02A}{`video1'}, hemispherical: \href{https://youtu.be/QCdHLPZE6k0}{`video2'}, elliptical: \href{https://youtu.be/v1UD0pAH_LM}{`video6'}).}}}
		\label{fig:6}
	\end{figure*}
	
	\subsubsection{Shadowgraph flow visualization} \label{shadow}
	The flow past spiked body configurations is visualized using the standard Z-type shadowgraph technique \citep{247182:5609367}. Shadowgraph imaging captures the density variations in the shock-laden flow field. A high-intensity plasma light (white light) source of 25 W is used to produce the required light intensity for shadowgraph imaging. The light is allowed passing through a slit which forms the point light source for shadowgraphy. Planar mirrors are used to deflect the light to the parabolic mirrors and the camera. A parallel beam of light is formed using the parabolic reflecting mirrors of 310 mm diameter and 2750 mm focal length to pass through the test section as shown in Figure \ref{fig:4}. A Phantom V211, monochrome, 12-bit high-speed camera is utilized to record the shadowgraph images at 256 $\times$ 160 pixels frame with a pixel resolution of 0.5 mm/pixel. The unsteady flow field around all the configurations is captured at a frame rate of 43000 Hz with the least exposure time of 2 \ensuremath{\mu }s. A maximum frame rate of 140740 Hz is achieved at a similar pixel resolution but at the reduced frame size of 128 $\times$ 64 pixels for exceptional cases of imaging. The refracted light is captured using a lens system in front of the camera. A 40 mm diameter achromatic doublet lens having an aperture of $f/2.5$ and a focal length of 100 mm is used to focus the light on the camera sensor. \sk{An effective total depth-of-field for the present imaging is calculated to be around 70 mm. The depth of field, the light intensity, and the optical arrangement are kept constant in the present study for qualitative comparisons.}
	
	Typical instantaneous shadowgraph images for different spiked body configurations are shown in Figure~\ref{fig:6}. Some of the critical flow features like weak leading edge shock, separation shock, shear layer, recirculation region, and reattachment shock could be identified from it. The flow field around the flat-face spiked body configuration (Figure~\ref{fig:6}a) stands out completely different from the flow fields that are observed around the hemispherical (Figure~\ref{fig:6}b) and elliptical (Figure~\ref{fig:6}c) spiked body configurations. \sk{Also, the qualitatively estimated shock strength from the shadowgraph images shows that the detached shock in-front of the flat-face spiked body configuration is the strongest. The strength of the shock was qualitatively evaluated by the thickness of the shock seen inside the red-dotted box, computed though an \textit{ad hoc} approach using an in-house Matlab edge detection algorithm} The shock strength in the other cases is found to decrease from the hemisphere to ellipse, as $\xi$ is decreasing (see the dotted red color boxes in Figure~\ref{fig:6}). The flow field around the hemispherical (Figure~\ref{fig:6}b) and elliptical (Figure~\ref{fig:6}c) spiked body configurations exhibit similarity in terms of overall flow features. However, they have varying shock strength, and slightly different flow morphology.

	\subsubsection{Drag force measurements} \label{drag} 
	An in-house built six components strain gauge balance has been used for the overall drag coefficient ($C_d$) measurements. A DC voltage of 18 V is used to excite the balance having a capacity to measure the axial and normal force of 50 kg and 250 kg, respectively. Only the axial force has been considered in the present investigation as the model is mounted at $\alpha=0^{\circ}$.
	
	\subsubsection{Steady and Unsteady pressure measurements} \label{press.meas.}
	Steady-state pressure measurements are carried out using the low response time Honeywell\textsuperscript{\textregistered}-Model FP2000 type configurable pressure transducers. The transducers measure the gauge pressure and have a response time of 2 ms. Two different range (sensitivity) of pressure transducers are used\textemdash50 psi (1000 $\mu$V/psi) and 100 psi (500 $\mu$V/psi). The data acquired from these sensors are sampled at 5 kHz with an accuracy of $\pm$0.1\% from the measured value. Pressure taps ($\sim$1 mm in diameter) are made at several locations (see Figure~\ref{fig:11}a) along the model forebody between $0.15<[S/D]<0.9$, and they are connected to the sensors using an integral polyurethane 1.5 m cable passing through the model interior.
	
	Unsteady pressure transducers (Kulite\textsuperscript{\textregistered}: XCL-100-50A of head size $\sim$2.6 mm), \sk{having a maximum frequency response of 250 kHz are used to measure the pressure fluctuations generated on the spiked forebody due to shock motions and the interaction of multi-scale turbulent structures}. \hlt{The locations of the unsteady pressure transducers are at $[S/D]=0.25$ for the flat-face, $[S/D]=0.4$ for the hemispherical, and $[S/D]=0.39$ for the elliptical forebody configurations (see Figure~\ref{fig:11}b-d)}. A sampling rate of 50 kHz for a total period of 2 seconds has been used for the final experiments after considering the effects of aliasing. The unsteady pressure transducers are flush-mounted near the reattachment shock corresponding to an azimuthal location of $\phi=90^\circ$.
	
	\subsubsection{Data acquisition system} \label{daq}
\begin{figure}
		\centering{\includegraphics[width=0.9\columnwidth]{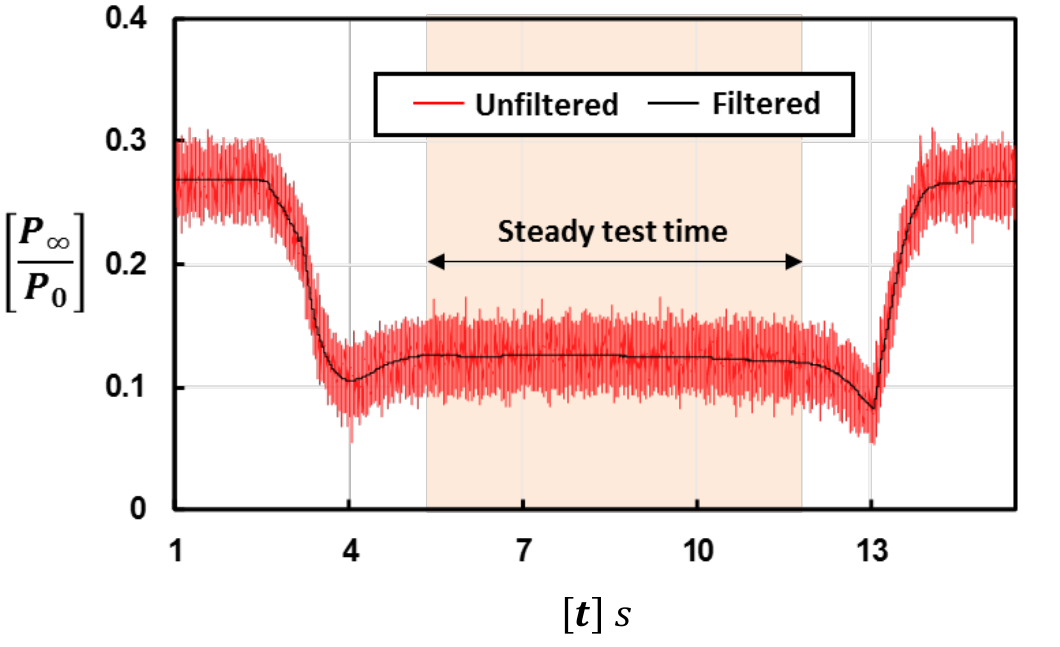}}
		\caption{\hlt{Typical static pressure signal (unfiltered \& filtered) observed during the supersonic wind tunnel operation at $M_\infty=2.0$. Location of the static pressure port is marked in Figure \ref{fig:4}}}
		\label{fig:7}
	\end{figure}

	The electrical output of all the measured parameters (forces, pressures, and angle of attack) is acquired using a National Instruments data acquisition system (NI-DAQ) having 32 sequential measuring channels. The sampling rate for all the analog channels is 250 kHz. The acquired signals are filtered, amplified, and transformed into digital data. The digital data are then transferred to a PC to evaluate the tunnel operating conditions, aerodynamic coefficients, and pressure coefficients. The obtained pressure histogram is further processed using a Matlab program to extract the power spectral density (PSD) and other useful unsteady statistics. The filtered and unfiltered static pressure data measured at the tunnel wall (see $P_\infty$ in Figure \ref{fig:4}) during a typical test is plotted in Figure~\ref{fig:7}. All the experimental data have been acquired during the steady test time, as indicated in Figure~\ref{fig:7}.
	
	\begin{table}
		\caption{\hlt{Flow conditions achieved in the test section during a typical run along with measurement uncertainties (given in percentage about the measured values).}}
		\bigskip
		\label{tab:2}
		\begin{tabular}{@{}ll@{}}
			\toprule
			\textbf{Quantities} &
			\textbf{Values}\\ 
			Total Pressure ($P_0$) &
			3.48 $\times 10^5$ \ensuremath{\pm} 5\% (Pa)\\
			Total Temperature ($T_0$) &
			294.46 \ensuremath{\pm} 2\% (K)\\
			Freestream Mach number ($M_\infty$) &
			2.01 \ensuremath{\pm} 1\%\\
			Freestream Temperature ($T_\infty $) &
			163.59 \ensuremath{\pm} 2\% (K)\\
			Freestream Pressure ($P_\infty$) &
			0.44 $\times 10^5$ \ensuremath{\pm} 5\% (Pa)\\
			Freestream Velocity ($U_\infty$) &
			515.15 \ensuremath{\pm} 2\% (m/s)\\
			Dynamic Viscosity ($\mu_\infty$) &
			$1.04 \times 10^{-5}$ \ensuremath{\pm} 2\% (Pa s)\\
			Freestream Density ($\rho_\infty $) &
			0.94 \ensuremath{\pm} 5\% (kg/m\ensuremath{^{3}})\\
			Reynolds number  &
			$2.16 \times 10^{6}$ \ensuremath{\pm} 5\%\\
			($Re_D=U_\infty D/\nu_\infty$, $D$ = 50 mm)&\\ \toprule
		\end{tabular}
	\end{table}  
	
	\subsection{Operating conditions} \label{op.con.}
	All the experiments are carried out at a freestream Mach number of $M_\infty=2.0$ with a settling chamber pressure of $P_0=3.5$ bar. The freestream Reynolds number based on the base diameter ($D$) is $Re_D=2.16\times 10^6$. Necessary wind tunnel calibration for the $M_\infty$ in the test-section has been conducted before the experiments. The freestream static pressure fluctuation intensity $\left(\sqrt{\overline{{\left(P_\infty-\bar{P}_\infty\right)}^2}}\right)$ is observed to be 1.57\%, which provides an estimate for the freestream turbulence. Table~\ref{tab:2} shows the achieved tunnel operating conditions during a typical run. 
	
	\begin{table}
		\caption{\hlt{Tabulation of different parameters including the drag coefficient ($c_d$) (with and without a spike), the pressure loading ($\zeta$), and the pressure fluctuation intensity ($\kappa$) observed for different spiked forebody configurations with varying geometric shape factor ($\xi$) at $M_\infty=2.0$.}}
		\bigskip
		\label{tab:3}
		\def\arraystretch{1}
		\centering
		\begin{tabular}{@{}lccccccc@{}}
			\toprule
			{\textbf{Cases}} & {$\boldsymbol{\left[S/D\right]}$} & \multicolumn{2}{c}{$\boldsymbol{c_d}$ ($\pm5\%$)} & \multicolumn{2}{c}{$\boldsymbol{\zeta}$ ($\pm5\%$)} & \multicolumn{2}{c}{$\boldsymbol{\kappa}$ ($\pm5\%$)} \\ 
			\hline
			&  & \textbf{no-spike} & \textbf{spike} & \textbf{no-spike} & \textbf{spike} & \textbf{no-spike} & \textbf{spike}  \\
			\hline
			ST-F\footnote{Sharp-tip spike on the flat-face forebody ($\xi \approx 1$)} & 0.25 & 1.51 & 1.06 & 5.85 & 4.84 & 0.025 & 0.452 \\
			ST-H\footnote{Sharp-tip spike on the hemispherical forebody ($\xi \approx 0.92$)} & 0.4 & 0.8 & 0.52 & 3.38 & 3.47 & 0.016 & 0.111 \\
			ST-E\footnote{Sharp-tip spike on the elliptical forebody ($\xi \approx 0.84$)} & 0.39 & 0.61 & 0.41 & 2.99 & 3.32 & 0.026 & 0.087 \\
			HA-H\footnote{Hemispherical aerodisk spike on the hemispherical forebody ($\xi \approx 0.92$)} & 0.4 & 0.8 & 0.35 & 3.38 & 2.09 & 0.016 & 0.107 \\ \toprule
		\end{tabular}
	\end{table}
	
	\section{Uncertainty} \label{uncert.}
	\hlt{Each experiment is repeated 5 times to ensure repeatability, and the reported data are the ensemble average of the 5 runs. However, built-in limitations of the instruments in terms of accuracy and precision in obtaining the measured quantities, bit rate conversions (analogue to digital, A/D), and the propagating errors in the derived quantities lead to inevitable uncertainties. Overall uncertainties in the experiments are calculated using the procedures described in the article of Moffat\cite{Moffat1988}, and in the book of Colemann and Steele\cite{Coleman2018}. In general, the known sources of uncertainty from the sensors, power supplies, A/D converters, and data acquisition cards are considered along with the measured uncertainty in image calibration, spatio-temporal resolutions and uncertainties in data-processing.} Uncertainty in the $c_d$ and unsteady pressure measurements is around $\pm5\% $. The uncertainties in the measurement of the other derived quantities are listed in Table~\ref{tab:2}. In the modal analysis (briefed in $\S$\ref{data_an} of the appendix), five sets of 1000 images are acquired from different runs, and they are processed individually to access the consistency in the final results. The reported values are the ensemble average from the five sets of modal analysis per case. The uncertainty in representing the spatial and temporal features are mainly dependent on the resolution itself (i.e., $0.5$ mm/pixel \& 43 Hz). Also, the spatial features are, in general, smeared due to the exposure limitations of the camera. Hence, there is an error associated with the exact representation of the spatial structure. However, for a 2 $\mu$s exposure, the smearing effect is found to be minimal, as reported in the work of \unskip~\cite{247182:5754299}. The recommendations of \unskip~\cite{247182:5754299} are followed to avoid aliased signals in the calculated spectral characteristics from the modal analysis. The total uncertainty involved in the spatial mode representation includes the error propagation in the image processing routines and the calibration procedure which is about $\pm4\% $. The total uncertainty in the temporal mode representation is around $\pm3\% $.
	
	\section{Results and Discussion} \label{res.disc.}
	
	\subsection{Drag reduction} \label{res.drag}
	Streamlined forebody offers lower drag than the blunt-body. The overall drag coefficient ($C_d$) is measured using the methods mentioned in $\S$\ref{drag}. However, for all practical purposes, the forebody drag coefficient ($c_d$) is preferred. Since the model is at $\alpha=0^\circ$, a pressure sensor kept in the base portion of the model is used to estimate the base drag coefficient, $C_{d,base}$ \citep{247182:8346719}. Finally, $c_d$ is calculated by subtracting $C_{d,base}$ from $C_d$. The values of $c_d$ for all the configurations are measured with and without a spike at $M_\infty=2.0$ and tabulated in Table~\ref{tab:3}. The elliptical spiked body configuration, which has the lowest $\xi$ provides the minimum $c_d$ in comparison to the hemispherical and flat-face spiked body configurations. Mounting of a spike, reduces $c_d$ significantly in all the configurations. Nevertheless, the spike, as is elaborated in the upcoming sections, causes severe flow unsteadiness.
	
	\begin{figure*}
		\centering{\includegraphics[width=\textwidth]{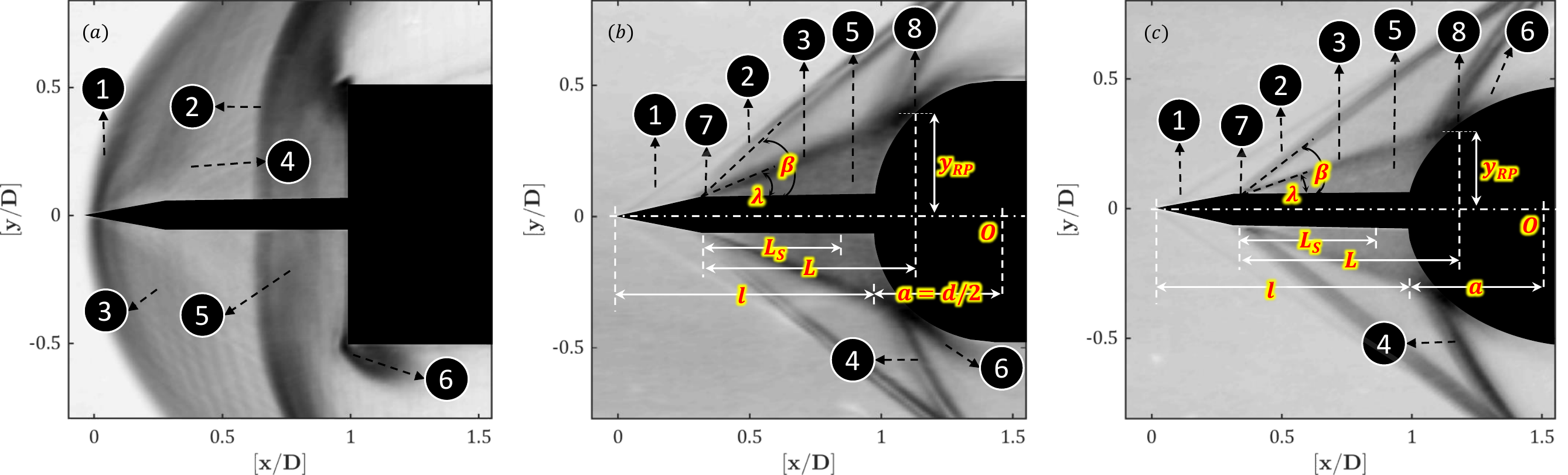}}
		\caption{\hlt{Time-averaged shadowgraph image obtained through the operation of $\left\|\bar{\boldsymbol R}-\boldsymbol R_{rms}\right\|$ for the (a) flat-face spiked body configuration, (b) hemispherical spiked body configuration, and (c) elliptical spiked body configuration at $M_\infty=2.0$, to highlight the spatial zones of dynamic flow events. Flow is from left to right. Flow features in (a): 1. Leading edge shock (S1), 2. Forebody shock (S2), 3. Trace of inflating shock (curved lines), 4. Trace of separation shock foot (oblique lines), 5. Torroidal vortex (TV) region. Flow features in (b) \& (c): 1. Weak leading edge shock, 2. Separation shock (SS), 3. Separated shear layer, 4. Reattachment shock (RS), 5. Recirculation region, 6. Trace of ejected fluid mass along the shear layer ($M_E$), 7. Separation point (SP), 8. Reattachment point (RP).}}
		\label{fig:10}
	\end{figure*}    
	
	\subsection{Unsteady pressure spectrum around the forebodies}\label{spectra} 
\begin{figure*}
		\centering{\includegraphics[width=0.7\textwidth]{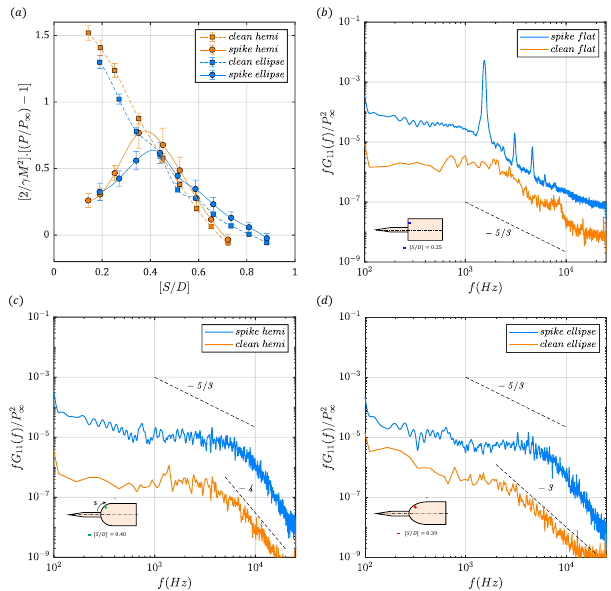}}
		\caption{\hlt{(a) Pressure coefficient variation along the forebody curvature length $[S/D]$ for the hemispherical and the elliptical forebody configurations (with and without sharp spike tip) obtained through the time-averaged wall-static pressure measurements. Power spectral density obtained from the unsteady pressure measurements for the different forebody configurations (with and without sharp spike tip): (b) Flat-face at $[S/D]=0.5$, (c) Hemisphere at $[S/D]=0.5$, and (d) Ellipse at $[S/D]=0.4$. The location where the unsteady pressure measurements are carried out are also shown in the small snippet in each of the graphs for the spiked body configurations at $M_\infty=2.0$.}}
		\label{fig:11}
	\end{figure*}

	A normalized operator-based time averaging ($\left\|\bar{\boldsymbol R}-\boldsymbol R_{rms}\right\|$, see $\S$\ref{data_an} in the appendix for explanation) is applied to the shadowgraph images in order to study the key dynamic features around the different spiked forebody configurations qualitatively as shown in Figure~\ref{fig:10}. The darker zones in the images mark the locations of dynamic flow events. In  Figure~\ref{fig:10}a associated with a flat-face spiked body, darker regions are observed, suggesting severe shock related unsteadiness. Figure~\ref{fig:10}b-c associated with the hemispherical and elliptical forebodies have different intensities of darkness, especially near the interface of the reattachment shock and the separated shear layer. Steady and unsteady pressure sensors are mounted near those regions to measure the level of shock-related unsteadiness, as shown in Figure~\ref{fig:11}. Locations and values of mean wall-static pressure coefficients ($c_p=[2/\gamma M^2]\cdot [(P/P_\infty)-1]$) for both the spiked and unspiked bodies are shown in Figure~\ref{fig:11}a, whereas the normalized unsteady pressure spectrum for each of the spiked configurations are shown in Figure~\ref{fig:11}b-d. The location of the unsteady pressure sensor is selected after identifying the maximum $c_p$ based on the results from Figure~\ref{fig:11}a for each of the configurations except the flat-face, as shown in the bottom-left portion of the schematic in Figure~\ref{fig:11}c-d. 
	
	A non-dimensional parameter termed as pressure loading ($\zeta $) is defined to quantify the net pressure acting on the respective forebody surface due to the shock-laden flow field. Similarly, the unsteady flow features near the reattachment shock of the forebody are represented through a parameter termed as pressure fluctuation intensity ($\kappa$) as shown in equations \ref{eq-kappa} and \ref{eq-zeta}, respectively, and also tabulated in Table~\ref{tab:3}. From the table, it can be seen that $\zeta $ is larger for the flat-face ($\zeta=4.84$) than the hemispherical ($\zeta=3.47$) and the elliptical ($\zeta=3.32$) counterpart. The flat-face spiked body configuration shows a maximum $\kappa$ of 45\%, whereas the hemispherical and elliptical spiked body configurations show only 11\% and 8.7\%, respectively. The values of $\zeta$ and $\kappa$ for each of the configurations with varying $\xi$ are also in agreement with the instantaneous shadowgraph images shown in Figure~\ref{fig:6} and also with the shocks and shear layers intensity shown in Figure~\ref{fig:10}.
	
	\begin{align}
		\label{eq-zeta}
		\zeta=\frac{P_{rms}}{P_\infty},\\
		\label{eq-kappa}
		\kappa=\frac{P_{s}}{P_{rms}}.
	\end{align}
	where,
	\begin{equation*}
		P_{rms}=\sqrt{\frac1n\sum_{i=1}^{n}P_i^{2}},\;P_{s}=\sqrt{\frac1n\sum_{i=1}^{n}({P_i-\overline P})^{2}},
	\end{equation*}
	\begin{equation*}
		P_i\;=\;\overline P+P',\;\overline P=\frac1n\sum_{i=1}^{n}P_i.
	\end{equation*}
	
	The existence of higher $\zeta$ and $\kappa$ for the flat-face spiked configuration can be explained through the previous works of \cite{247182:5562226} and \cite{247182:5562227}. The pulsation mode of unsteadiness is the primary reason for the observation. The pulsation mode consists of three phases termed as `collapse', `inflation', and `withhold' \citep{247182:5562227}. These three phases collectively cause the entire shock system ahead of the flat-face spiked body to move forth and back periodically, as shown in Figure~\ref{fig:6}a, and \ref{fig:10}a. Hence, the unsteady pressure spectrum shows a dominant peak at 1338.3 Hz ($St=fU_\infty/D=0.13$) with harmonics (see Figure \ref{fig:11}b). The Strouhal number ($St$) is in-fact a bit lower than the value of 0.183 from the empirical relation (see equation \ref{eq-flat}) given by \cite{247182:5562226} for the same $(l/D)$.
	
	\begin{equation}
	\label{eq-flat}
	S_t=0.25-0.067\left(\frac lD\right).
	\end{equation}
	
	However, in the cases of hemispherical and elliptical spiked forebodies, the shock oscillations are localized at the points of separation and reattachment of the shear layer, as shown in Figure \ref{fig:10}b-c, which result in lower $\zeta$ and $\kappa$. The unsteady pressure spectra in these two cases have no discrete frequencies but only broadband (see Figure \ref{fig:11}c-d). In Figure \ref{fig:11}c, the power spectrum reveals a broadband spectrum between 2000-7000 Hz (blue line) in comparison to the no-spike case (orange line). \hlt{This range of amplified frequencies is most likely associated with the shear layer instabilities which form large scale structures (LSS) carrying shocklets that later impinge on the forebody itself. A similar broadband spectrum is also observed in the same frequency range for the elliptical case but with a slightly lower amplitude, as shown in Figure \ref{fig:11}d.}
	
	\subsection{Shock related unsteadiness in the spiked bodies} \label{rd3}
	\subsubsection{Shock pulsation in the flat-face spiked body configuration} \label{case1}
	
\begin{figure*}[!ht]
		\centering
		{\includegraphics[width=0.9\textwidth]{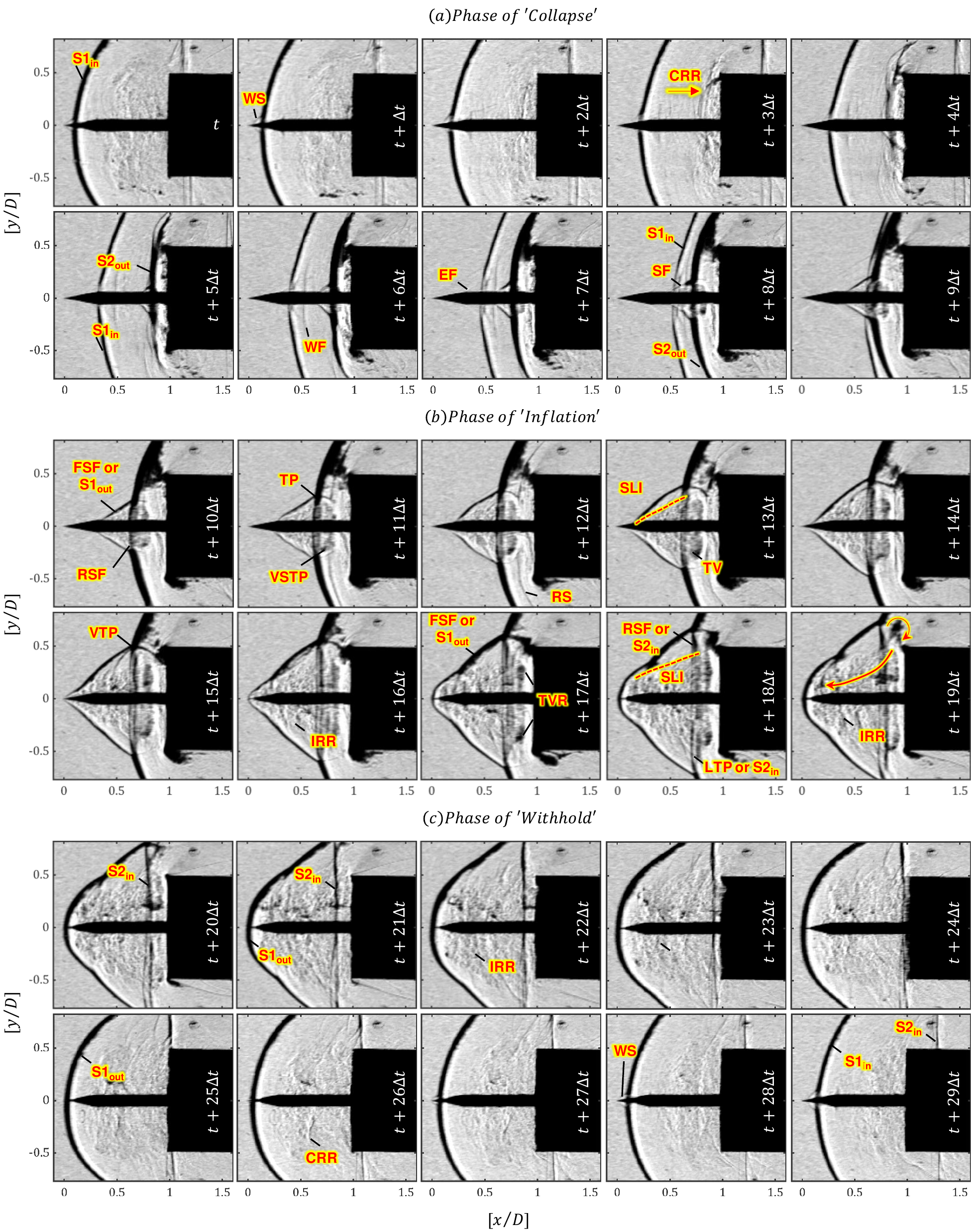}}
		\caption{\label{fig:13}Instantaneous shadowgraph images at different time intervals showing the detail features of (a) collapse, (b) inflation and (c) withhold during the pulsation mode of unsteadiness over a spiked flat-face forebody configuration at $M_\infty=2.0$. Flow is from left to right. Dominant flow features: $S1_{in}$-shock from the leading edge of the spike moving towards the body, WS-weak leading edge shock, $S2_{out}$-shock moving away from the forebody, WF-weak compression front, EF-expansion fan, SF-shock foot, FSF or $S1_{out}$-front shock foot, RSF or $S2_{in}$-rear shock foot, SLI-shear layer interface, TP-triple point of SF, VSTP-vortex shedding from the triple point of SF, RS-refracted shock, TV-torroidal vortex formed from the TP, VTP-vertically moving TP, IRR-inflating recirculation region, TVR-torroidal vortex ring, LTP-locus of TP and trace of RSF about the axis, CRR-convecting recirculation region along the flow direction. Corresponding video file \href{https://youtu.be/nYG-BfhB02A}{(video1)} is available in the supplementary.}
	\end{figure*}

	Figure~\ref{fig:10}a shows the time-averaged flow field obtained from the sequence of instantaneous shadowgraph images. The two extreme positions of the pulsating shock, along with the weak leading-edge shock, are observed. A complete cycle of the shock pulsation including all the three different phases (collapse, inflation, and withhold), as mentioned by \unskip~\cite{247182:5562226} and \unskip~\cite{247182:5562227} are represented in the instantaneous shadowgraph images in Figure~\ref{fig:13}. In the image, the timestamps for the representative instantaneous shadowgraph images are indicated in which $\triangle$t$=1/f_s$, where $f_s$ is the sampling rate of the camera. Unless otherwise noted, $f_s$=43000 Hz for all the configurations. The `phase of collapse' is captured between t to t+9$\triangle$t, where the shock system approaches towards the forebody (Figure~\ref{fig:13}a). Between t+12$\triangle$t to t+18$\triangle$t, the `phase of inflation' is captured. In this phase, the two shock systems are seen to merge and start inflating radially upstream (Figure~\ref{fig:13}b). Finally, between t+21$\triangle$t to t+26$\triangle$t, the `phase of withhold' is observed, where the inflated shock stands at the tip of the spike before it starts to collapse again to repeat the cycle (Figure~\ref{fig:13}c). As the pulsation mode has been discussed by the other researchers \cite{247182:5562226,247182:5562227}, a brief description regarding Figure \ref{fig:13} is given in $\S$\ref{flow_flat-face}.   
	
	\begin{figure*}
		\centering
		{\includegraphics[width=0.8\textwidth]{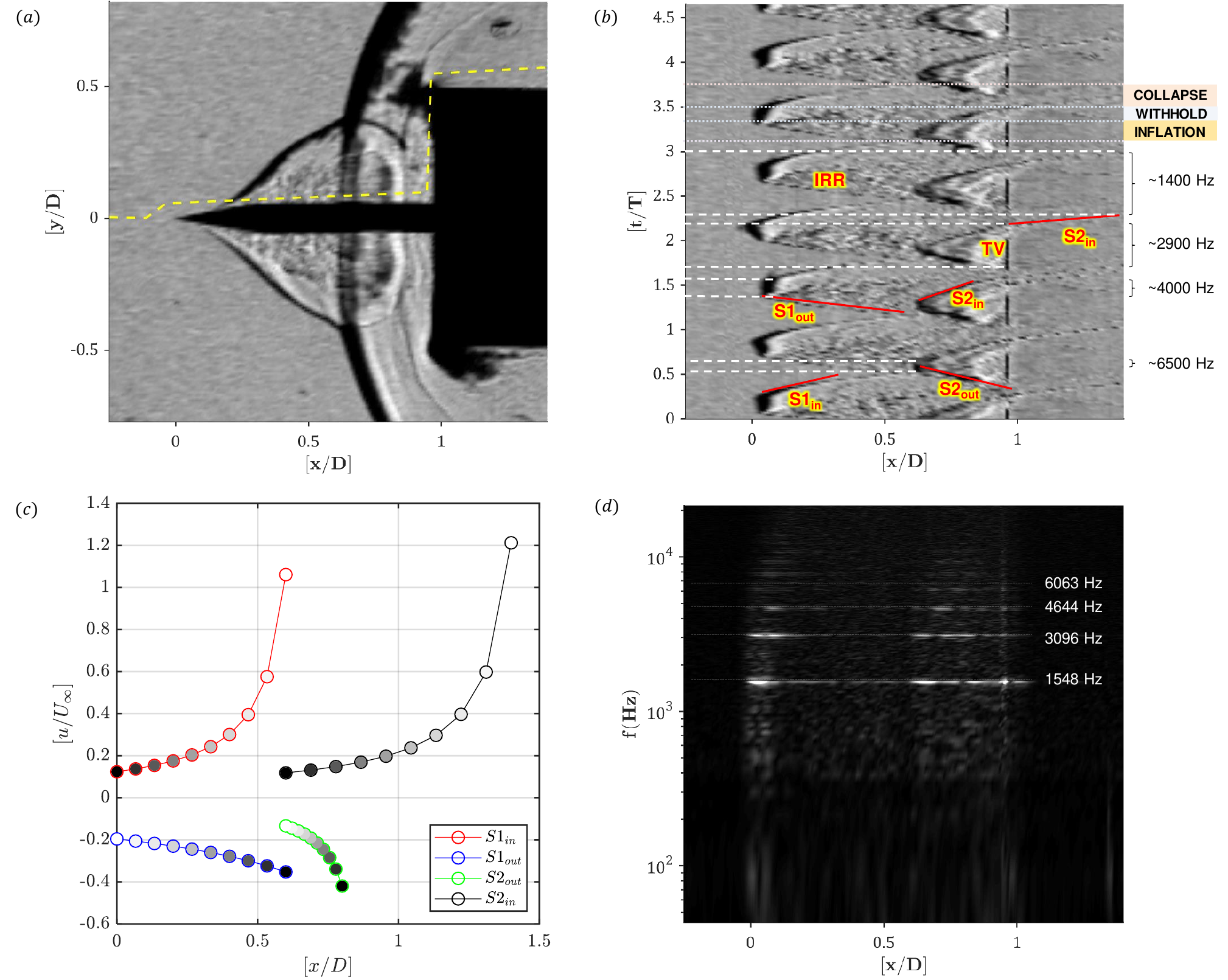}}
		\caption{\hlt{Simple sectional image analysis routines performed for the flat-face spiked body configuration at $M_\infty=2.0$. (a) The dashed yellow line represents the considered section in a typical instantaneous image; (b) Time evolution (normalized with $T=1$ ms) of the sectional intensity scan revealing the dominant features in the flow \sk{(the mentioned approximate frequencies are from the visual assessment of the xt-diagram indicating the expected time scales at different stages of oscillation);} (c) $x-t$ trajectory analysis performed to compute the velocity of the shock motion - The direction of time is represented through the shades of the filled circles from dark to bright; (d) Normalized spectral analysis of the sectional images evolving in time revealing the dominant frequency contents in the flow. Flow is from left to right.}}
		\label{fig:15}
	\end{figure*}
	
	A $x-t$ diagram shown in Figure\ref{fig:15} is constructed by stacking the line segments of the instantaneous shadowgraph images. It is essentially a temporal footprint of the dominant flow structures (in our case, it is the shock). In the $x-t$ diagram (Figure \ref{fig:15}b), the three events of the pulsation cycle are marked based on the explanations mentioned in $\S$\ref{flow_flat-face}. In the `inflation' zone, the trace of $S1_{out}$ is seen. At `withhold,' $S1_{out}$ could be seen to stay at a constant place for some time, and during `collapse,' $S1_{in}$ is seen to accelerate (Figure \ref{fig:15}c) towards the body (Figure \ref{fig:15}b). In the supplementary, a video (\href{https://youtu.be/nYG-BfhB02A}{`video1'}) of the high-speed shadowgraphy depicting the pulsation mode of unsteadiness is given for reference.
	
\begin{figure*}
		{\includegraphics[width=\textwidth]{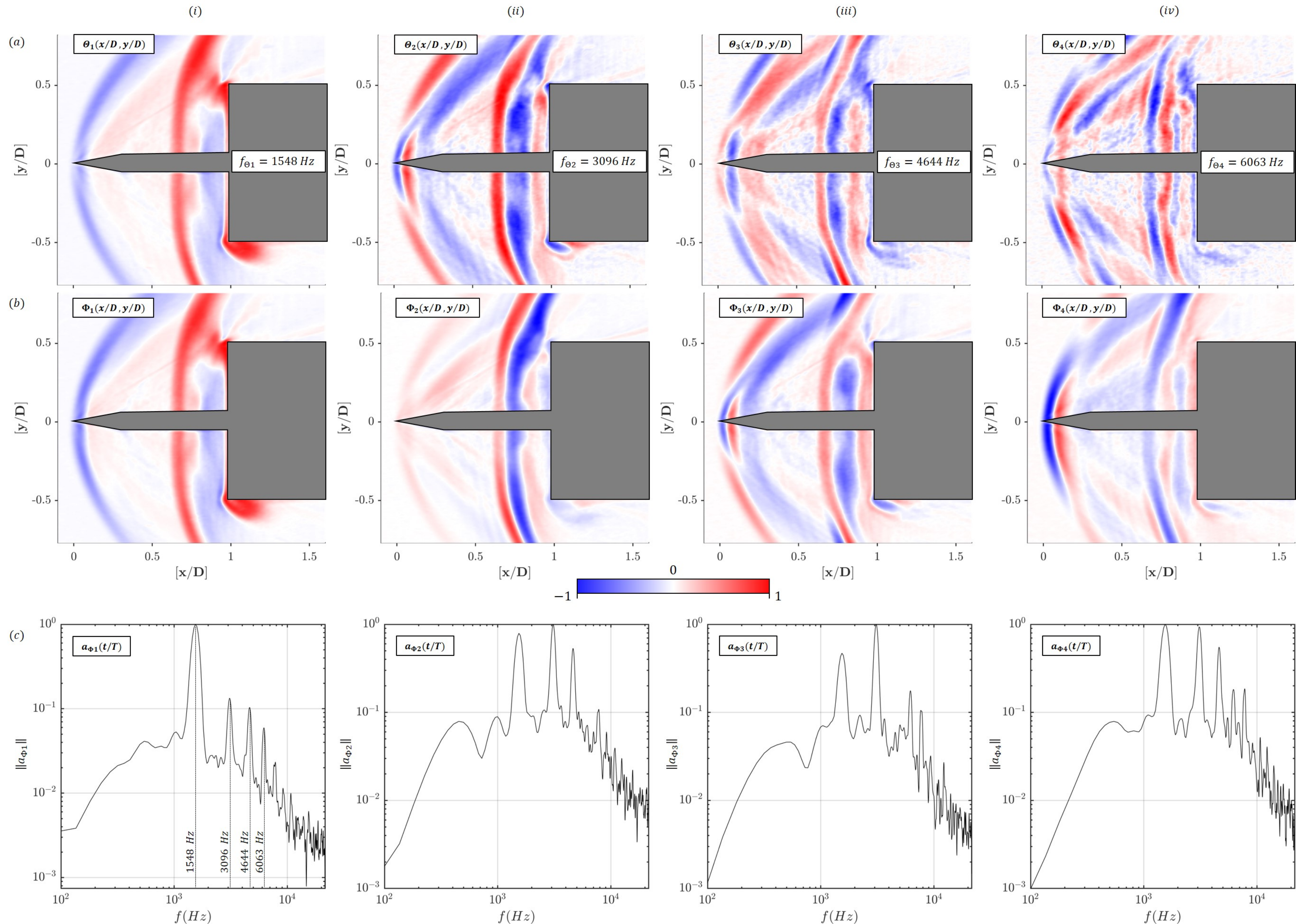}}
		\caption{\sk{Comparison of the first four dominant (a) dynamic and (b) energetic spatial modes obtained from the DMD (${\Theta_{1-4}(x/D,y/D)}$) and the POD (${\Phi_{1-4}(x/D,y/D)}$) analysis for the flat spiked body configuration at $M_\infty=2.0$. Flow is from left to right (Contour plots of the spatial modes from the modal decomposition mark the positive (red) and negative (blue) correlation throughout the manuscript). (c) Spectral analysis of the POD temporal coefficients ($a_{\Phi_{1-4}}(t)$) for ${\Phi_{1-4}(x/D,y/D)}$ (The spectra from the DMD analysis are given in Figure~\ref{fig:9}c to avoid cluttering).}}
		\label{fig:16}
	\end{figure*}

	Modal decomposition is performed on the obtained shadowgraph images using the procedure mention in Appendix $\S$\ref{data_an}. The dominant dynamic [${\Theta_{1}(x/D,y/D)}$] and energetic [${\Phi_{1}(x/D,y/D)}$] spatial modes are observed to be the same as shown in Figure \ref{fig:16}a-(i) and \ref{fig:16}b-(i). The dominant spatial modes have persistent spatial structures that are well correlated with the time averaged shadowgraph image (see Figure~\ref{fig:10}a). Such correlation is expected as the dynamic and energetic contents are extracted from the single time lagged matrix (see equations \ref{pod4}-\ref{pod5}), by performing single value decomposition (see equation \ref{pod8}) and eigenvalue decomposition (see equation \ref{pod10}). In addition, the temporal coefficients of the dominant POD mode ($a_{\Phi1}(t/T)$), reconstructed from the temporal signal of the DMD analysis ($a_{\Theta}(t/T)$), and the normalized unsteady pressure signal for the flat-face spiked body configuration ($P/P_\infty$) are also shown to be closely matching in Figure \ref{fig:9}d. Thus, the spatiotemporal modes from the modal analysis are verified and validated. 
	
	From Figure~\ref{fig:16}a-b, the spatial modes from DMD and POD analyses are very similar, except little grains in DMD spatial modes. The dynamic contents are sensitive to the spatial resolution and the image noise. Hence, the DMD spatial modes given in Figure~\ref{fig:16}a are grainy. However, POD spatial modes (Figure~\ref{fig:16}b) are energy based orthogonal projections. Hence, they are not sensitive to such temporal noises. For further discussions, only energetic spatial modes are used. The first spatial mode ($\Phi_1(x/D,y/D)$), represents the time-averaged shadowgraph image (see Figure~\ref{fig:10}a). The dominant temporal content ($a_{\Phi_1}(t/T)$) for $\Phi_1(x/D,y/D)$ is observed at a fundamental frequency ($f_f[a_{\Phi1}(t/T)]$) of 1548 Hz from Figure~\ref{fig:16}c-(i). There are at least three harmonics present ($f_1[a_{\Phi1}(t/T)]=3096\;Hz,\;f_2[a_{\Phi1}(t/T)]=4644\;Hz,\;f_3[a_{\Phi1}(t/T)]=6063\;Hz$), however, they have lower amplitudes. Also, the unsteady pressure spectra obtained at $[S/D]$=0.25 for the flat-face case in Figure~\ref{fig:11}b compares well with the above findings.
	
	The analysis of the first four energetic spatial modes ($\Phi_1(x/D,y/D)$ to $\Phi_4(x/D,y/D)$) gives information about the positions of the shocks and their progression over a given time period (Figure~\ref{fig:16}). The modal analysis gives the correlation parameters based on the shock motion. The red and blue contour levels provide positive and negative correlations. By observing the color levels of $\Phi_1(x/D,y/D)$, it can be seen that $S1$ and $S2$ are moving exactly in opposite directions. The actual direction could not be ascertained using a single mode, but the sense of relative direction between the shocks can always be identified. The first and last snapshot of the collapse phase in Figure \ref{fig:13}a, $S1$, and $S2$ are observed at their extreme forward locations ($S1_{out}$ and $S2_{out}$). The spectra of the POD coefficients in Figure \ref{fig:16}c-(i) contain the same frequencies as observed from the unsteady pressure measurements (Figure \ref{fig:11}b). Thus the dominant energetic spatial mode could be inferred to represent the complete cycle itself. Both DMD (Figure \ref{fig:9}c) and POD (Figure \ref{fig:16}c-(i)) spectra, confirm the existing of a dominant spectral peak around 1550 Hz. In Figure \ref{fig:15}d, from the $x-t$ diagram, the dominant event is also observed to be around 1550 Hz. While looking at Figure \ref{fig:15}b, the corresponding spectral line on the $x-t$ diagram is also found to be existing between the beginning and ending of a single pulsation cycle. Thus the dominant spatial mode $\Phi_1(x/D,y/D)$ from POD analysis is identified to be pulsating.
	
	In the second dominant energetic spatial mode ($\Phi_2(x/D,y/D)$), the terminal phase of `collapse' is represented (Figure \ref{fig:16}b-(ii)). The shock $S1$ has refracted through $S2$, and it might be approaching the flat-faced forebody. While looking at the temporal content, the first harmonic ($f_1[a_{\Phi2}(t/T)]=3096\;Hz$) is reaching an amplitude equivalent to that of the fundamental ($f_f[a_{\Phi2}(t/T)]=1548\;Hz$). It can be seen from the $x-t$ diagram (Figure \ref{fig:15}) that the shock $S2_{out}$ moves away from the body at a low speed, while the approaching shock $S1_{in}$ comes at a higher velocity. These two shocks interact and refract at about $[x/D]\approx0.7$, which is captured as the second dominant mode in Figure \ref{fig:16}b-(ii). The shock refractions are attributed to the popping up of the first harmonic ($f_1$) at twice the dominant frequency ($f_f$) in Figure \ref{fig:16}c-(ii). \unskip~\cite{247182:10453042} had also reported the occurrence of similar harmonics; however, it was not reasoned out clearly.
	
	Similarly, the third dominant energetic spatial mode ($\Phi_3(x/D,y/D)$) picks up a stage during the phase of `inflation' with torroidal vortex (TV) formation and the inflation of $S1_{out}$ shock (Figure \ref{fig:16}b-(iii)). In this case, the temporal mode yields a dominant frequency, which is the first harmonic ($f_1[a_{\Phi3}(t/T)]=3096\;Hz $) as shown in Figure \ref{fig:16}c-(iii). It makes sense, as the refracted shock yields the formation of a toroidal vortex system, while the $S2_{in}$ shock completely changes its structure into a lambda shock system.
	
	The fourth energetic spatial mode ($\Phi_4(x/D,y/D)$), represent the beginning and the terminal phase of `withhold' (Figure \ref{fig:16}b-(iv)). The shock is standing at the tip of the spike ($S1$) and the temporal mode ($a_{\Phi4}(t/T)$) yields a spectral content equivalent of the first dominant energetic spatial mode ($\Phi_1(x/D,y/D)$) as shown in Figure \ref{fig:16}c-(iv). The other energetic spatiotemporal modes, carry only minimal fluctuations about the mean. Also, from the energy contents given in Figure~\ref{fig:9}a-b, it can be seen that the first four modes carry most of the energy contents ($>$30\%). Hence, we consider them to be sufficient to represent the entire flow field.
	
	Thus, the present analysis confirms most of the findings from the previous investigation by \unskip~\cite{247182:5562227} through a $x-t$ diagram and provided some more insights about the unsteady spatiotemporal modes. Consequently, the present investigation methodology can be used in the next cases of actual interests, to access the flow physics associated with varying shock related unsteadiness.
	
	\subsubsection{Strong shock interactions in hemispherical spiked body configuration} \label{case2}
	
	Looking at the spectra of Figure~\ref{fig:11}c, we notice that the discrete frequencies associated with the flat-face spiked forebody (Figure~\ref{fig:11}b) have vanished for the hemispherical spiked forebody configuration under the same freestream flow conditions. This observation is consistent with the previous reports of \unskip~\cite{247182:5562223} and \unskip~\cite{247182:5562224}. However, the underlying reason behind the broadband spectra of Figure~\ref{fig:11}c has not been clearly explained.
	
\begin{figure}
		\centering
		{\includegraphics[width=\columnwidth]{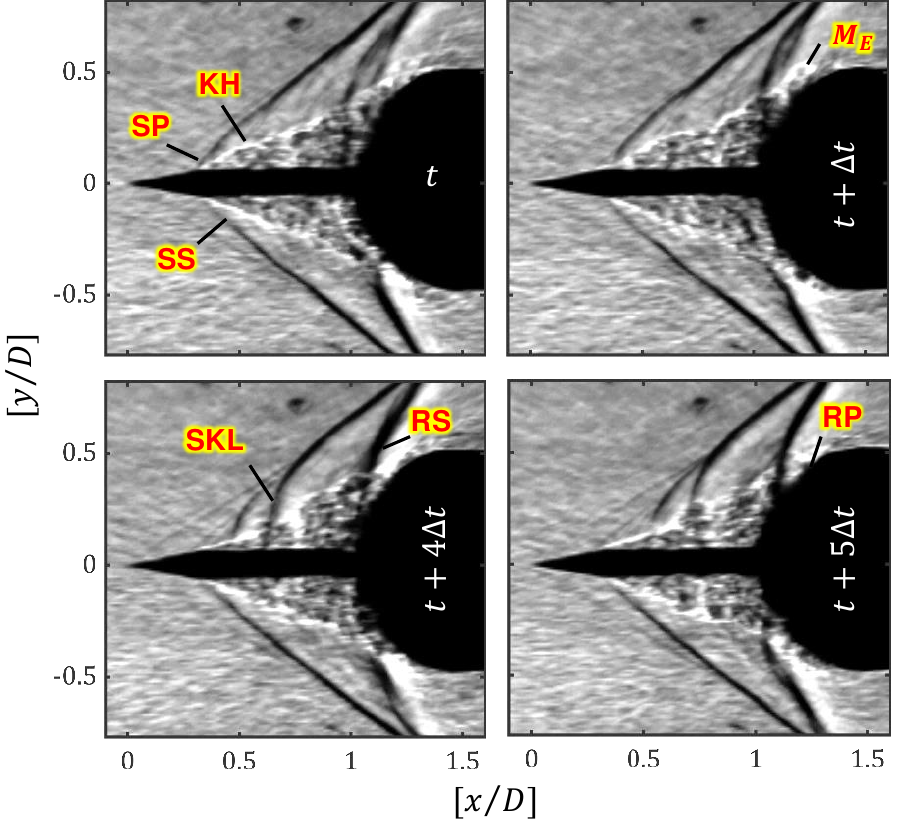}}
		\caption{\hlt{Instantaneous shadowgraph images at different time intervals showing the unsteady motion of the shock wave in a spiked hemispherical body configuration at $M_\infty=2.0$. Flow is from left to right. Some dominant features: SS - Separation shock, RS - Reattachment shock, SP - Separation point, KH - Kelvin-Helmholtz structures, SKL - Shocklets, RP - Reattachment point, $M_E$ - Ejected fluid mass. (Corresponding \href{https://youtu.be/QCdHLPZE6k0}{video} file is available in the supplementary)}}
		\label{fig:17}
	\end{figure}

	In order to explain the accompanied reduction of the pressure fluctuation intensity ($\kappa$) (Table \ref{tab:3}), time-resolved shadowgraph images are analyzed qualitatively in Figure~\ref{fig:17}. The flow history represented through instantaneous snapshots indicates the presence of shock-related unsteadiness, especially at the point of shock separation (SP) on the spike stem and the reattachment point (RP) near the forebody shoulder. \hlt{From the time-averaged shadowgraph images given in Figure \ref{fig:10}b, vital kinematic information like shock angle ($\beta$), flow deflection angle $\theta$, shear layer growth rate ($\delta'$) and local flow/convective Mach number ($M$ or $M_C$) can be estimated. The values are listed in the tabulation shown in Figure \ref{fig:23}, which will be helpful in further discussion. Growth rates $\delta'=\delta/x$ are calculated from the time-averaged images using an in-house Matlab code to detect the transverse extent of the visual shear layer thickness ($\delta$). A gradient function is utilized to identify the shear layer (medium-gray zone, Figure \ref{fig:10}), as the freestream is almost light-gray in color everywhere in Zone-3 and darker-gray everywhere in Zone-4 (Figure\ref{fig:23} to identify zones). The calculated values are found to be less than the ones reported by \citep{Slessor2000} in relation to the compressible shear layer growth rate. The reasons are attributed to the leading edge separation and to the fact that the separated shear layer is not exactly planar and partially bounded by walls.}
	
	\sk{The evolution of large scale structures (KH instabilities) from the SP and the movement of the RS (Figure~\ref{fig:17}) resemble the periodic charging and ejection of fluid mass from the recirculation region. The charging happens along the separated shear layer and the ejection can be seen as density blobs escaping at the end of the separated shear layer in the form of large scale structures or eddies. The ejected blobs were visualized in the schlieren images of a simple forward-facing step \citep{247182:5719032} in a supersonic flow. They are also shown in the present time-resolved shadowgraph images in Figure \ref{fig:19}-\ref{fig:20} (which will be discussed later). In Ref. \citep{247182:5719032}, the model under study resembles a flat-face spiked-body configuration. However, their spike is thicker and longer in comparison to the present study. Their protrusion is merely a flat-face forebody with a comparable base body diameter as that of their spike. Hence, the underlying flow features are equivalent, except that our forebody shape is hemispherical.} 
	
	\sk{In fact, based on the forward-facing step study, the ejection rate could be compared to the KH frequency through a relation $f_{ej} \sim 10^2f_0 \sim f_{KH}$, where $f_0$ is the dominant low-frequency unsteadiness. It is also similar to the shock-wave turbulent boundary layer interaction (SWTBLI) problems such as those induced by compression ramps, reflected shocks, protrusions, and fins \citep{247182:6199877}. However, the equivalency presented in \citep{247182:5719032} comes only from the assumption that the low-frequency unsteadiness ($f_0$) scale effectively as the ratio of `mass ejection’ to `reversed mass flow rate’. Hence, factors like mean flow organization, interaction length/time scales, and shear layer characteristics are bound to play an important role in the aforementioned relation. Besides, the equivalency is only true for the well-developed separation as seen in the cases of protrusion or swept separations \citep{247182:5719032,Huang2018}. In low-deflection configurations like the streamlined forebodies (hemispherical or elliptical in the present study), such an event is not expected to be exactly the same and hence, the scaling relations should be approached with caution.} 
	
	In any case, knowledge of these canonical forms of SWTBLI flow problems gives better insights to the present analysis. These events are highly non-linear and broadband in nature due to multi-scales involved in the unsteady process, which further explains the broadband signal seen in Figure \ref{fig:11}c. In addition, the majority of the above-said SWTBLI problems arise due to either the upstream influence \citep{247182:5925124,247182:5925118,247182:5925119} or the downstream influence \citep{247182:5925175,247182:5925123}. For the upstream influence to be dominant, the incoming boundary layer should be of a considerable thickness to influence the oscillation of the separation shock foot. The downstream influence is attributed to the recirculation bubble dynamics, which drive the oscillation of both the separation and reattachment shocks (see Figure 1 of \unskip~\cite{247182:6199877}).
	
	Very recently, in the paper of \unskip~\cite{247182:5719032}, the authors have shown that in the problem of SWTBLI interactions, the free shear layer drives the shock-related unsteadiness. They have shown that the volume inside the recirculation region increases once the recirculation bubble is charged with the freestream fluid. Therefore, it achieves a pressure value that is sufficiently larger than the static pressure existing behind the RS (reattachment shock). At this point, the authors have shown, using their unsteady pressure measurement data, that the RS is pushed upstream of the body, and the SS is pushed towards the body. At this moment, the excess fluid mass inside the recirculation bubble is reportedly ejected to the freestream. However, from the range of protrusion heights used in their experiments, they have concluded that the free shear layer thickness and the separation length ($L_S$) increase with the protrusion height. It has been shown that the free shear layer interaction drives the shock-related unsteadiness. Also, in their study, the authors realized that the separation angle ($\lambda$) associated with $L_S$ does not change with protrusion height.

	\begin{figure*}
		\centering
		{\includegraphics[width=0.9\textwidth]{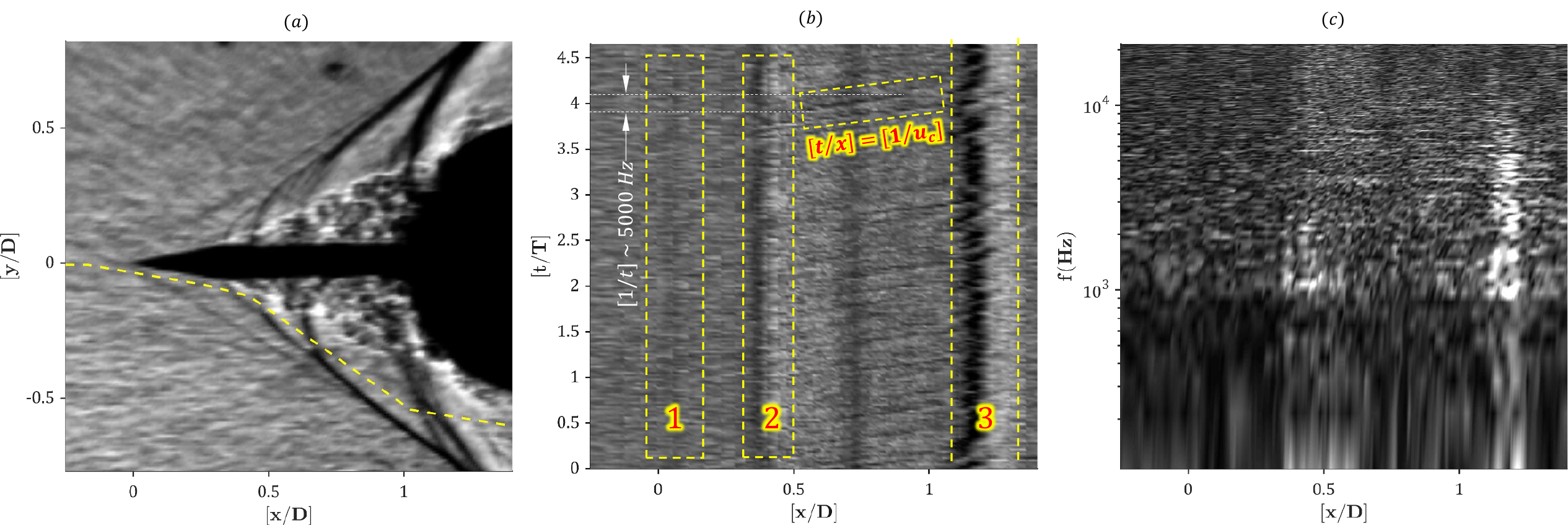}}
		\caption{{Simple sectional image analysis routines performed for the hemispherical spiked body configuration with a sharp spike-tip at $M_\infty=2.0$. (a) Yellow-line represents the considered section in a typical instantaneous image; (b) Time evolution (normalized with $T=1$ ms) of the sectional intensity scan revealing the dominant features in the flow \textemdash 1. Weak leading-edge shock in-front of the sharp spike-tip, 2. Separated shock, 3. Oscillating reattachment shock near the hemispherical forebody; (c) Normalized spectral analysis of the sectional images evolving in time ($x-t$ diagram) revealing the dominant frequency contents in the flow. Flow is from left to right.}}
		\label{fig:18}
	\end{figure*}  
	
	In our present study, the boundary layer thickness upstream of the SP over the spike stem, is very thin for $l/D=1$. Consequently, it does not provide any significant upstream influence on the shock dynamics. Hence, the only phenomena that might be driving the shock-related unsteadiness are the downstream influence. As shown in Figure~\ref{fig:17}, a detached shock pulse or shocklet (SKL) is seen to be generated near the SS. These SKLs propagate downstream along with the large scale (KH) structures from SP to RP and interact with the reattachment shock (RS), severely. As the large scale structures convect along with the separated shear layer from the SP to the RP, they also grow in size. These structures transport the mass, momentum, and energy of the flow from the freestream to the recirculation zone and form a visible axisymmetric cone around the spike stem. 
	
	The SKLs are carried along with the large scale structures (as seen in Figure \ref{fig:17}) just like the wavefronts carried by the structures in the convecting free shear layer of a supersonic jet \citep{247182:8346725,247182:8346726}. The feet of SKLs are inside the separation bubble, and they cannot be seen in the present images due to the line of sight integration. However, it can be seen from the video provided in the supplementary (\href{https://youtu.be/EEVu-MS4-X8}{`video3'}), which is based on our accompanied numerical studies\footnote{The DES results from the numerical studies are given in a qualitative manner here, just to appreciate the observations from the present analysis. Details about the computation methods are beyond the scope of the current paper and hence, it is not discussed at this point. Some vital information about the numerical studies are given in the video caption itself.}.  When the large scale structures along with the SKLs hit the RS, the foot of the SKL hits the forebody and reflects back and forth. When another SKL comes in contact with the wall or the reflected SKL from the wall, the velocity of the reflecting and refracting SKLs changes drastically. At this point, the large scale structures also hit the RS and thus, displacing the shock slightly. During such an event, the charged fluid mass from the recirculation bubble is ejected ($M_E$) to the freestream. In the meantime, reflected waves from the forebody wall further propagate upstream towards the SP. While doing so, the reflected wave interacts with the incoming SKLs and perturbs the SP. A perturbed SP produces a large scale structure that gets amplified then due to the KH instabilities present in the free shear layer. Thus, the unsteadiness is maintained.
	
	\begin{figure}
		\centering
		\includegraphics [width=\columnwidth] {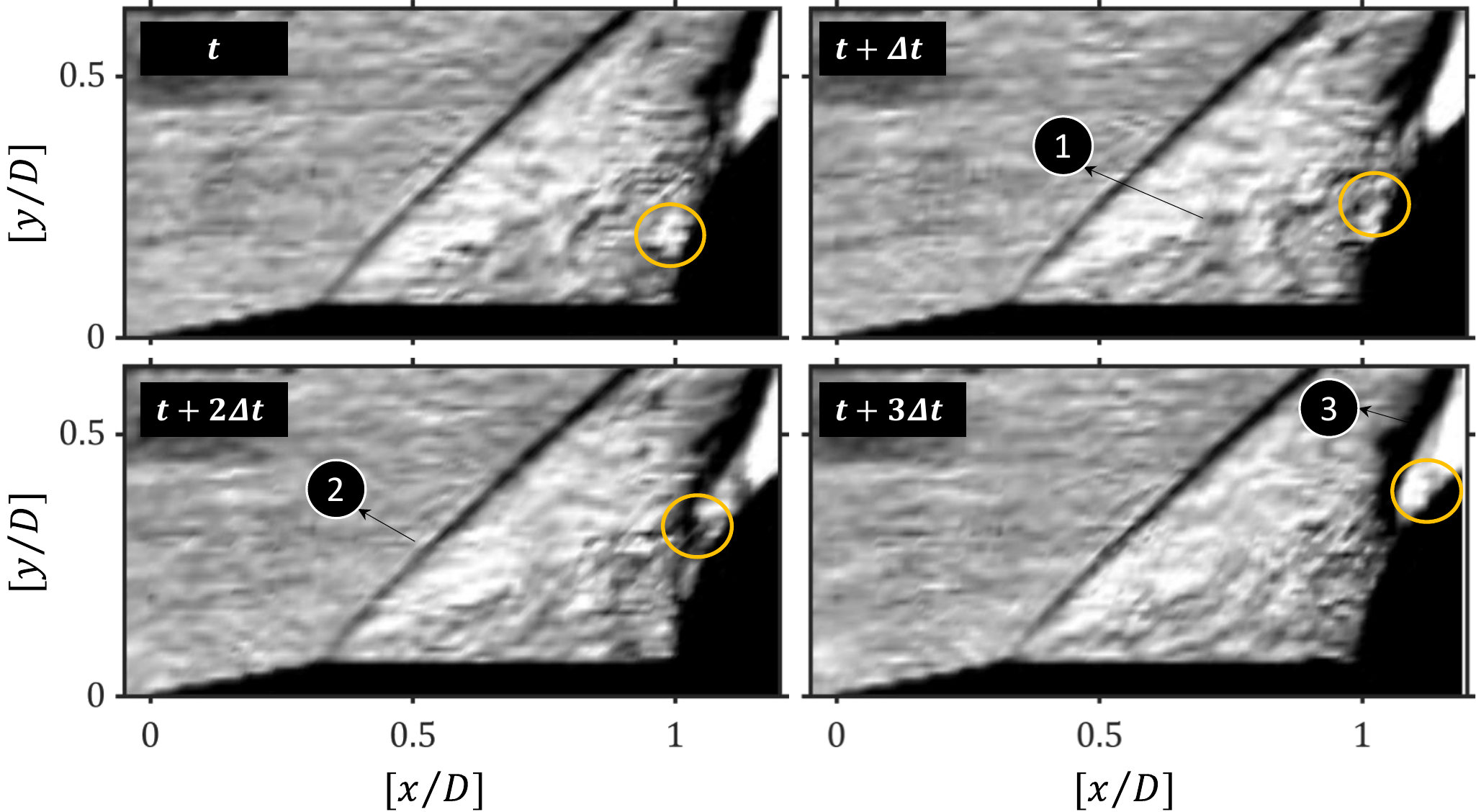}
		\caption{\hlt{Instantaneous shadowgraph images at different time intervals captured at a frame rate of $f_s=140470$ Hz showing mass ejection across the oscillating reattachment shock of a spiked hemispherical body configuration at $M_\infty=2.0$. Golden-yellow circle is shown to track the movement of a fluid mass ejecting out. Flow is from left to right. Dominant flow features: 1. Kelvin-Helmholtz (KH) structures, 2. Separation shock, 3. Reattachment shock. (Corresponding video file is available in the supplementary under the name \href{https://youtu.be/2LAdCpIQAP8}{`video4'}, and \href{https://youtu.be/gHpjJcdnB58}{`video5'})}}
		\label{fig:19}
	\end{figure}
	
	A simple $x-t$ diagram reveals many of the unsteady events happening in the hemispherical spiked body configuration (Figure \ref{fig:18}). The line along which the shock footprints are obtained with respect to time is marked in Figure \ref{fig:18}a. The shock footprints are shown in Figure \ref{fig:18}b and the corresponding spectral analysis is given in Figure \ref{fig:18}c. Weak leading edge shock, SS, and RS are seen in Figure \ref{fig:18}b. The motion of SKLs, along with the large scale structures, are seen as inclined lines with a slope of [$1/u_c$]. The slope of the line directly provides the convective velocity ($u_c$), and information about the frequencies of the SKLs carried along with the large scale structures. Also, at the SP, the fluctuation of the shock footprint is minimal. Whereas, it becomes more substantial near the RP (region-3 marked with yellow dashed lines in Figure \ref{fig:18}b). The reason is due to the impingement of the growing large scale structures near the RS. Performing spectral analysis on the shock footprint reveals the presence of broadband spectra near the SP and RP (Figure \ref{fig:18}c).
	
	\begin{figure}
		\centering
		{\includegraphics [width=\columnwidth] {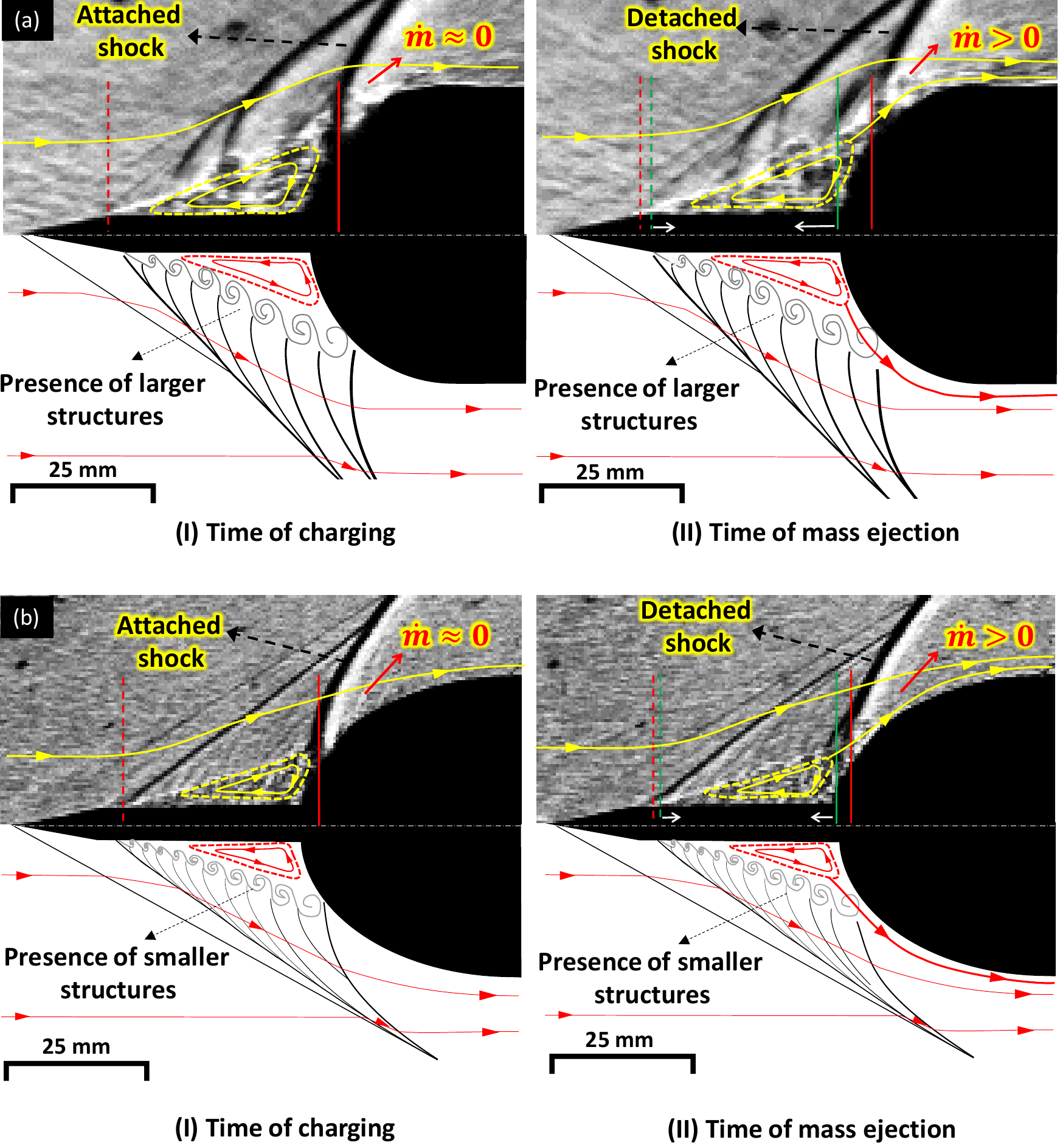}}
		\caption{\hlt{Instantaneous shadowgraph snapshots showing the motion of the separation and reattachment shocks during the time of charging and ejection for the (a) hemispherical and (b) elliptical spiked body configurations at $M_\infty=2.0$. Flow is from left to right. The mirrored images about the axis contain the sketches and annotations of the observed flow events. Yellow/red solid line with arrows - streamline; Yellow/red dashed line - recirculation zone; Vertical red/green dashed line - initial/final locations of the separation shock during the time of charging; Vertical red/green solid line - initial/final locations of the reattachment shock during the time of ejection.}}
		\label{fig:20}
	\end{figure}
	
	The charging and ejection of fluid mass are also seen in the recirculation region represented in the sequence of shadowgraph images captured at a higher frame rate ($f_s=140740$ Hz), as shown in Figure~\ref{fig:19}. A small semi-transparent golden-yellow circle is marked around a fluid blob that is initially inside the recirculation region. As time progresses, the fluid blob is ejected out into the freestream flow. The RS is pushed ahead of the forebody, and the SS is slightly moved towards the body in the opposite direction (see also \cite{247182:5719032}). Also, the large-scale structures produced along with the shear layer due to the KH instability are seen in the second frame of Figure~\ref{fig:19}. Findings from the $x-t$ plots given in Figure \ref{fig:18} are also consistent with the above statements. Videos showing these phenomena at a frame rate of both $f_s=43000$ (\href{https://youtu.be/QCdHLPZE6k0}{`video2'}) Hz and $f_s=140740$ (\href{https://youtu.be/gHpjJcdnB58}{`video5'}) Hz are given for reference in the supplementary. 
	
	\hlt{In the forward-facing step research \citep{247182:5719032}, the frequency of the shock foot movement is reported, especially in the separation and reattachment zones on the axisymmetric-cylinder but not on the protrusion itself. A dominant range of low frequencies ($10^2-10^3$ Hz) is seen at the point of separation, and a dominant range of high frequencies ($\sim 10^4$ Hz) is seen at point of reattachment. In our study, however, it was impossible to mount the pressure transducer at the shock foot location as the spike stem is very thin. Hence, we have measured the pressure spectra at the forebody shoulder (reattachment shock location) which is mainly governed by the shear layer unsteadiness (Figure \ref{fig:11}c).}
	
	\hlt{From Figure~\ref{fig:19}, in the t+$\Delta$t snapshot, one could see the KH instability, clearly. The wavelength is identified to be about $[x/D]\sim 0.15$. From the $\theta-\beta-M$ analysis provided in the table of Figure \ref{fig:23}, the convective velocity is estimated as $\sim 317$ m/s. Using these values, the KH instabilities characteristic frequency is identified to be around $\sim 10^4$ Hz, which is the approximate upper limit of the amplified range of frequencies. However, the sensor head itself is having a diameter of $[d_s/D]\sim 0.1$. Hence, the spectral resolution of the KH instabilities corresponding to the $\sim 10^4$ Hz would be difficult. In \citep{247182:5719032}, the authors used $D=75$ mm, and a sensor of $[d_s/D]\sim 0.01$, which lead them to finely resolve the spectrum from the KH instabilities $\sim 10^4$ Hz. Moreover, as they increase the ratio of $h/\delta$ from $1.0$ to $5.0$ (where $h$ is the protrusion height and $\delta$ being the boundary layer thickness at the point of separation), the dominant higher frequency contents shift towards lower frequency, from $\sim 3 \times 10^4$ Hz to $\sim 1 \times 10^4$ Hz while the range of amplified spectral contents is getting broader. In this regard, in our case, the equivalent ratio is $h/\delta \sim D/2\delta \sim 25$, meaning that the frequency is expected to be lower and the range of frequencies being amplified wider.}
	
	\hlt{The broadband frequency seen in the unsteady pressure spectra of Figure \ref{fig:11}c is in $10^3-10^4$ Hz range. The frequency contents are arising from the shocklets convecting along the KH-instabilities and later impinge around the reattachment point. These interactions happen at multi-scales which are supported by the instantaneous shadowgraph images and the \href{https://youtu.be/EEVu-MS4-X8}{DES video} given in the supplementary. Considering the separation length ($L_S$), or the perimeter of the recirculation region as the characteristic length, and the associated velocity of the convective disturbances inside it, the derived approximated frequency will fall in-between the above mentioned range. The velocity scale of the disturbances along the shear layer, and the local sound velocity at which the disturbances are traveling inside the recirculation region are assumed to be approximately equivalent to the convective velocity ($[u/u_c]\sim [a/u_c]\sim 1$). The time scale of the disturbance in the recirculation region or the eddy turn-over time is calculated by taking the characteristic length $([l^*/D] \sim [L_S/D] \sim [l/D]\sim 1)$. Thus, the frequency contents are computed to be $[u_c/l^*]\sim 10^3$ Hz. Since, the shocklets ride on the KH-instabilities, the spectral contents could be associated with the shear layer instabilities.}
	
	The movement of the separation and reattachment shocks in the opposite phase during the charging-ejection sequence is shown more clearly in Figure~\ref{fig:20}a. The shocks system are seen to be moving in the opposite phase. The dashed and solid vertical lines mark the positions of the separation and reattachment shocks on the forebody, respectively. The red and green colors of the vertical dashed/solid lines marks the initial and final location of the respective shocks during the time of charging (Figure~\ref{fig:20}a-I) and ejection (Figure~\ref{fig:20}a-II). At the time of charging, the reattachment shock is attached to the forebody, and during mass ejection, it is observed to be detached.
	
	\begin{figure}
		{\includegraphics[width=\columnwidth]{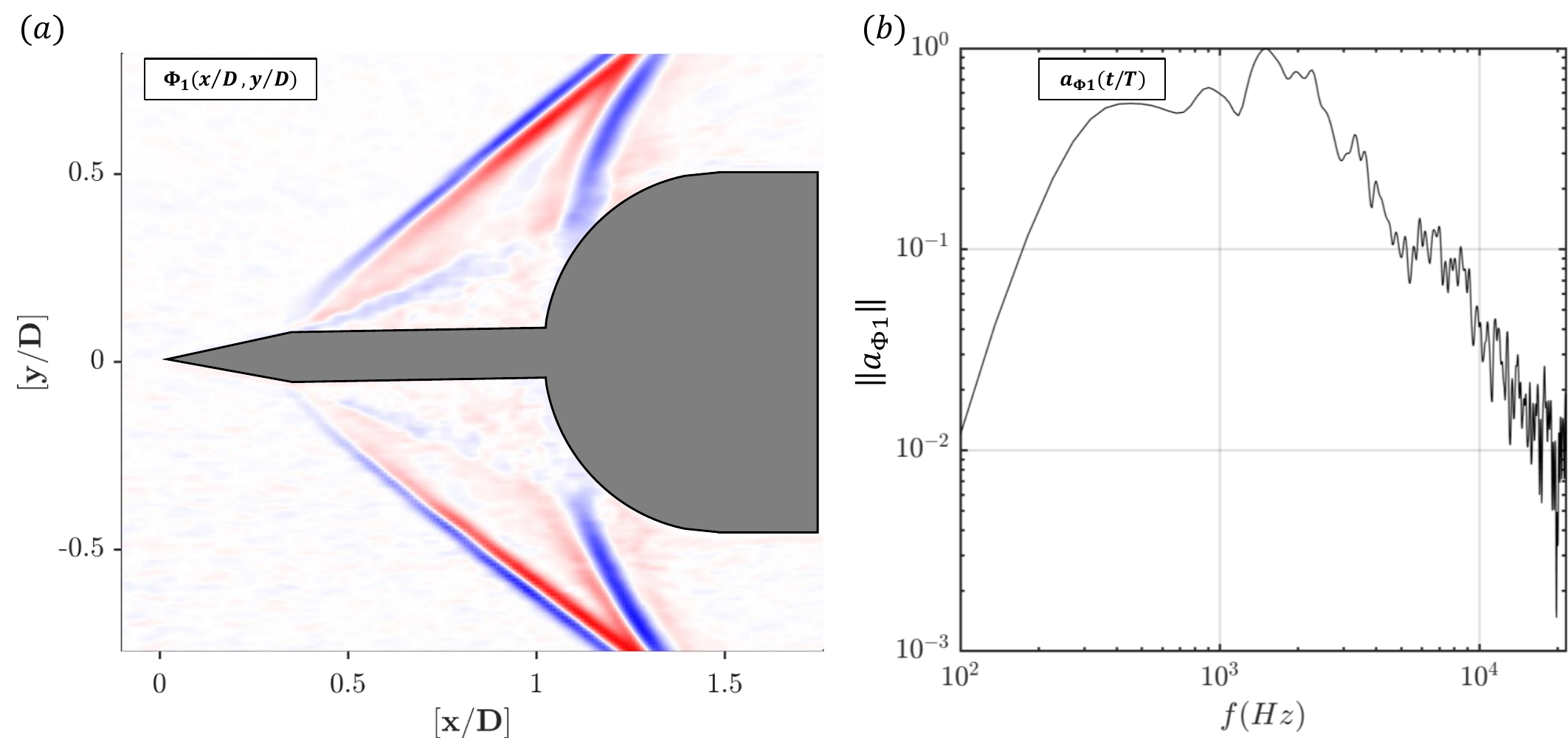}}
		\caption{{(a) Dominant energetic spatial mode ($\Phi_1(x/D,y/D)$) and (b) the normalized spectrum of the temporal mode ($a_{\Phi1}(t/T)$) obtained from the POD analysis of the shadowgraph images for the hemispherical spiked body configuration $M_\infty=2.0$. Flow is from left to right.}}
		\label{fig:21}
	\end{figure}
	
	From the POD analysis, the dominant spatial mode ($\Phi_1(x/D,y/D)$) and the corresponding temporal coefficient ($a_{\Phi_1}(t/T)$) spectrum are extracted as shown in Figure \ref{fig:21}. In Figure~\ref{fig:21}a, $\Phi_1\left(x/D,y/D\right)$ shows the oscillations of the separation and reattachment shocks are out-of-phase (note the alternating color contours of the separation shock (blue-red), and the reattachment shock (red-blue)). A clear representation is given later in Figure \ref{fig:27}a, where the magnitude of such out-of-phase motion is marked. The observation of shocks system moving in opposite phase from $\Phi_1(x/D,y/D)$, in turn, supports the existence of recirculation bubble dynamics with charging and ejection mechanism as proposed by \unskip~\cite{247182:5719032} in their studies on the axisymmetric protuberance at supersonic speeds. From Figure \ref{fig:9}a-b in $\S$\ref{data_an}, it is also shown that there is no single dominant mode. From Figure \ref{fig:9}b, it is observed that the cumulative energy contained in each of the modes does not vary linearly (on a $\log$ scale) for the hemispherical case, unlike the flat-face spiked configuration. Hence, it requires at least $\sim$70 modes to represent $\sim$30\% of the energy in the flow, unlike $\sim$4 modes for the flat-face case. The presence of a higher number of spatial modes to represent 30\% of energy contents emphasizes the presence of turbulent structures at different scales (broadband).
	
	\sk{As a cautionary note, it has to be written that the POD/DMD modes from the shadowgraph images can only be used to identify the large scale movements like the shocks but clearly not the motion of smaller eddies. The results from the model decomposition tools utilizing the shadowgraph images are only used to highlight the presence of out-of-phase shock motions and the multi-scale events. However, intrinsic features like the ejection frequencies from the mass transporting eddies may not be extracted with sufficient clarity given the line-of-sight light integration and the shallow depth-of-field.} 
	
	The spectral characteristics from $a_{\Phi_1}(t/T)$ (see Figure \ref{fig:21}b) is also consistent with the findings from the unsteady pressure measurement in Figure~\ref{fig:11}c (blue color line showing pressure spectra at S/D=0.4) and also with the dynamic contents computed from the DMD analysis (see orange colour line in Figure~\ref{fig:9}b). In Figure~\ref{fig:11}c, the shear layer frequencies are about 4000-7000 Hz. However, in Figure~\ref{fig:9}b, the peak is at 1000 Hz. This is because we consider the entire spatial mode instead of a point measurement. Asides $\Phi_1(x/D,y/D)$, the other spatial modes represent the deviations from the asymmetry of the flow separation about the spike stem axis. Asymmetry is unavoidable due to the non-linear nature of the shocks system and the presence of helical modes precising about the axis, as seen in the previous work of \unskip~\cite{247182:5562223}. However, $\Phi_1(x/D,y/D)$ is not having any asymmetric effect because the dominant spatial mode is driven purely by the axial shocks motion.
	
	\begin{figure}
		\centering
		{\includegraphics[width=\columnwidth]{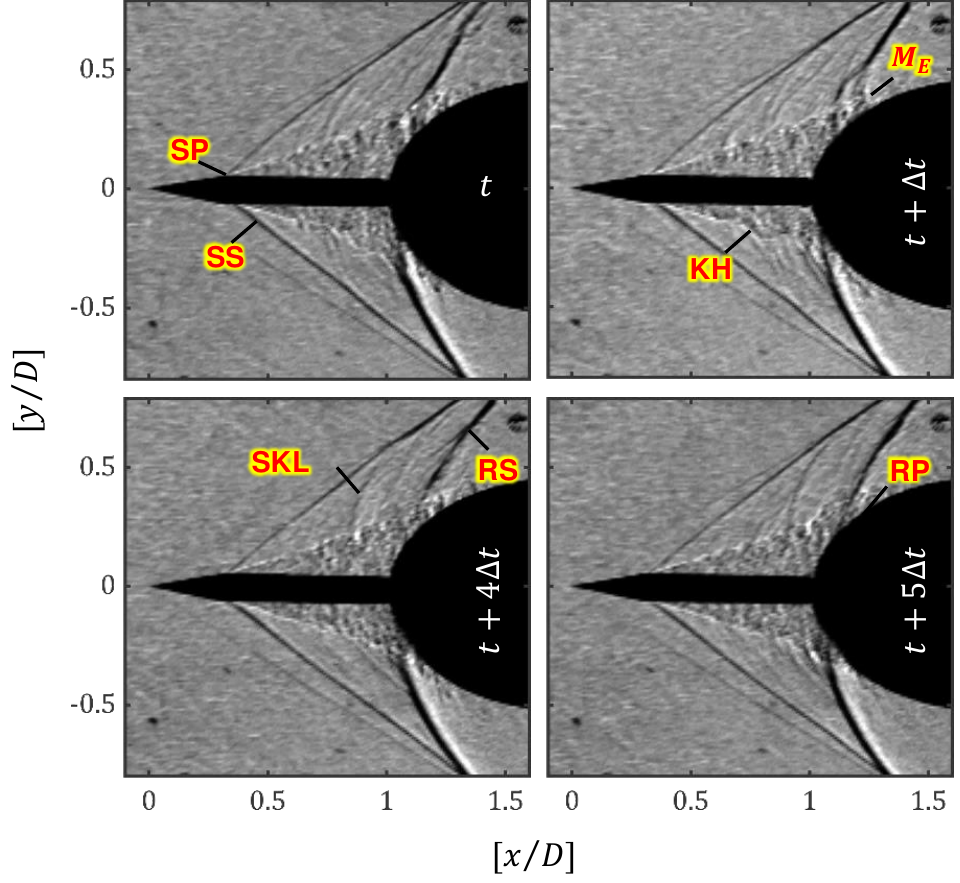}}
		\caption{\hlt{Instantaneous shadowgraph images at different times showing the unsteady shock motion in a spiked elliptical body configuration at $M_\infty=2.0$. Dominant features: SS - Separation shock, SP - Separation point, SKL - Shocklets, KH - Kelvin-Helmholtz structures, RP - Reattachment point, RS - Reattachment shock, $M_E$ - Ejected fluid mass. Flow is from left to right. (Corresponding \href{https://youtu.be/v1UD0pAH_LM}{video} is available in the supplementary)}}
		\label{fig:22}
	\end{figure}
	
	\subsubsection {Weak shock interactions in elliptical spiked body configuration} \label{case3}
	
	\begin{figure*}
		{\includegraphics[width=0.8\textwidth]{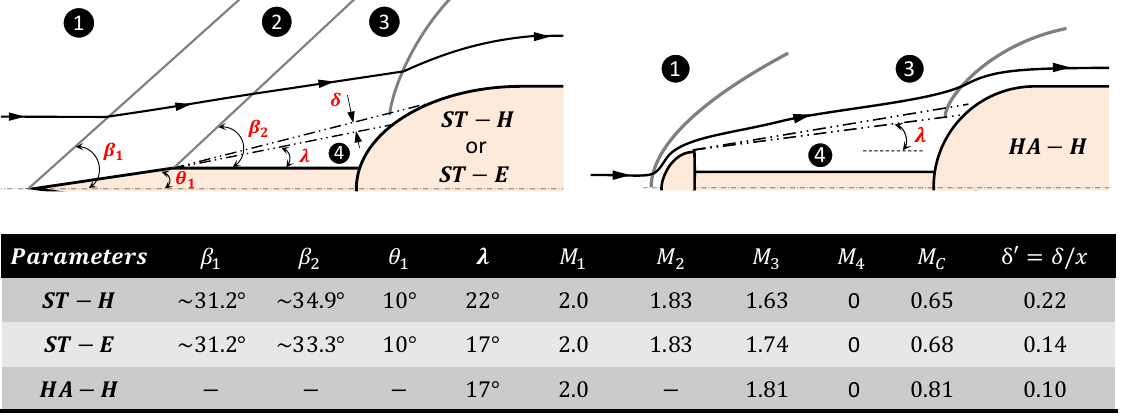}}
		\caption{\hlt{Schematic showing the different parameters and their respective values like the deflection angle ($\theta$), shock angle ($\beta$) and Mach number ($M$) in certain regions of the flow as calculated from the gas dynamic tables for an axisymmetric cone \citep{247182:5941997} at $M_\infty=2.0$. Flow is from left to right. Flow region: 1. freestream, 2. upstream shock separation, 3. downstream shock separation, 4. recirculation region. ST-H and ST-E: Sharp-tip spike on the hemispherical and elliptical forebodies, respectively. HA-H: Hemispherical aerodisk mounted on the hemispherical forebody.}}
		\label{fig:23}
	\end{figure*} 
	
	In this section, when the forebody shape ($\xi$) changes from hemispherical to elliptical, the reason behind the damping of shock-related unsteadiness is investigated. In the instantaneous shadowgraph image series shown in Figure~\ref{fig:22}, the separation point (SP), weakly oscillating separation shock (SS), weak shocklets (SKLs), free shear layer with large scale structures (KH instabilities), reattachment point (RP), and the weakly oscillating reattachment shock (RS) are seen. The term weak is used in comparison to the observations made in case of the spiked hemispherical forebody using shadowgraphy. In the supplementary, a high-speed shadowgraphy video is given for reference (\href{https://youtu.be/v1UD0pAH_LM}{`video6'}).
	
	The time-averaged shadowgraph image in Figure \ref{fig:10}c show that the location of RP is observed at a different height ($y_{RP}$) for each of the cases with different reattachment length ($L$). However, the separation length ($L_S$) is almost the same for both cases. In the elliptical case, RP shifts further downstream and approach closer to the axis, thereby increasing $L$ and reduce $y_{RP}$, in comparison to the hemispherical one. The values of $y_{RP}$ for both cases are also in agreement with the empirical relationship given by \unskip~\cite{247182:5939825} (see equation 18 therein). 
	
	\hlt{Relative to the hemispherical case, in the elliptical one, $L$ is increased, whereas the cone angle ($\lambda=17^\circ$) of the recirculation zone is decreased. Reduction in $\lambda$ leads to a lower separation shock angle ($\beta_2$) and consequently to a higher Mach number ($M_3=1.74$) behind the separation shock as shown in Figure \ref{fig:23}. In the hemispherical case, for the same flow conditions, $\lambda=22^\circ$ (higher) and $M_3=1.63$ (lower). All the relevant variables in Figure \ref{fig:23} are calculated using the simple gas dynamic relationships for supersonic flow ($\theta-\beta-M$) past an axisymmetric cone at $\alpha=0^\circ$ \unskip~\citep{247182:5941997}. }
	
	In an attempt to clarify the difference between the unsteadiness of hemispherical and elliptical spiked forebodies, the air velocity is approximated as zero in the recirculation zone (marked as region 4 in Figure~\ref{fig:23}). The convective Mach number ($M_C$) is obtained by following the procedure prescribed in \citep{Slessor2000}. For the elliptical case, $M_{C}$ is higher due to a larger $M_{3}$ in comparison to the hemispherical one. The growth rate of a compressible free shear layer thickness ($\delta'$) was also shown to increase when $M_C$ decreases \unskip~\citep{Slessor2000}. Therefore, in the case of the elliptical spiked body, $\delta'$ and the size of the roller structures are comparatively lower. Thus, $\lambda $, $L$, and $M_3$, are the important parameters influencing the intensity of shock-related unsteadiness in both configurations.
	
	In Figure~\ref{fig:20}b, the separation shock foot motion and the extremes of the reattachment shock oscillation are given qualitatively for the elliptical spiked body configuration. The shock strength in the elliptical case (Figure~\ref{fig:20}b) is found to be weaker in comparison with the hemispherical one (Figure~\ref{fig:20}a). Looking at the intensity of the shock motion in both cases, as shown in Figure~\ref{fig:20}, the unsteadiness is observed to be damped considerably in the elliptical case. As the overall flow features between the hemispherical and the elliptical cases are similar, charging and ejection of fluid mass from the recirculation region is also seen in the present case. However, from the schematics drawn in Figure \ref{fig:20} and the video given in the supplementary (\href{https://youtu.be/gHpjJcdnB58}{`video6'}) for the elliptical case, the size of the large scale structures are smaller in the free shear layer, and therefore, are expected to have higher frequency contents.  
	
	\begin{figure*}
		\centering
		{\includegraphics[width=0.9\textwidth]{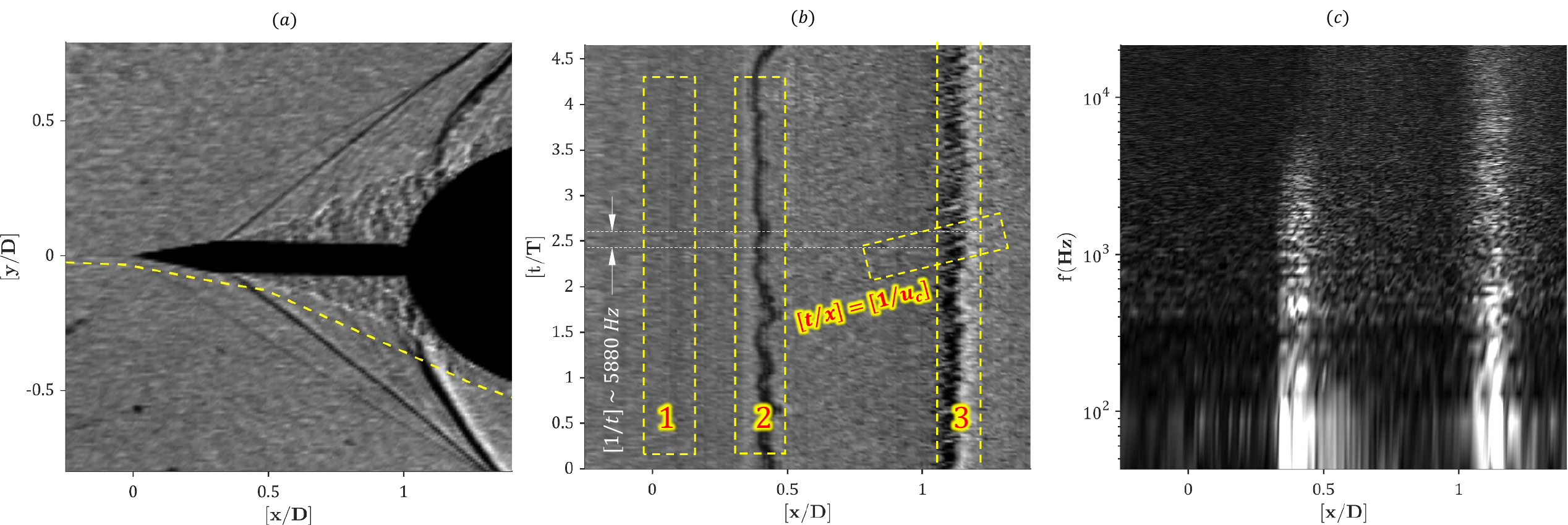}}
		\caption{\hlt{Simple sectional image analysis routines performed for the elliptical spiked body configuration with a sharp spike-tip at $M_\infty=2.0$. (a) Yellow-line represents the considered section in a typical instantaneous image; (b) Time evolution (normalized with $T=1$ ms) of the sectional intensity scan revealing the dominant features in the flow \textemdash 1. Weak leading-edge shock in-front of the sharp spike-tip, 2. Separated shock, 3. Oscillating reattachment shock near the elliptical forebody; (c) Normalized spectral analysis of the sectional images evolving in time ($x-t$ diagram) revealing the dominant frequency contents in the flow. Flow is from left to right.}}
		\label{fig:25}
	\end{figure*}

	A simple shock footprint analysis using the $x-t$ diagram, as shown in Figure \ref{fig:25}, sheds more light on the flow physics in the elliptical case. The temporal footprint of the weak leading edge shock, SS, and RS, are shown in Figure \ref{fig:25}b. As shown in the previous case (Figure \ref{fig:18}b), SKLs are also present in the current case between $[x/D] = 0.5$ and $1.0$. However, they are weak, and hence, a distinct trace of the SKLs in the elliptical case is not present. Weak SKLs in the elliptical case result in weaker reflection from the model forebody. Hence, they exert only smaller perturbations to the incoming SKLs and SP. Smaller disturbances in SP are not amplified as much as in the hemispherical case, which leads to the production of relatively finer structures. The size of these structures comparatively grows at a lower rate compared to the hemispherical case. Because they are smaller, when they make contact with the forebody at the RP, weaker oscillations of the RS are seen. Hence, near the reattachment zone (region-3) marked in Figure \ref{fig:25}b, the relative unsteadiness of the RS footprint is confined only to a distance of $[x/D] \sim 0.15$ (the distance between the yellow dashed vertical lines) for the elliptical case. For the hemispherical case, the RS footprint fills up the space of $[x/D] \sim 0.25$ (Figure \ref{fig:18}b). The corresponding spectrum given in Figure \ref{fig:25}c is a broadband one, just like the hemispherical case near the reattachment point. The spectrum of the unsteady pressure measurements shown in Figure~\ref{fig:11}d is also in agreement with the present findings.
	
	\begin{figure}
		{\includegraphics[width=\columnwidth]{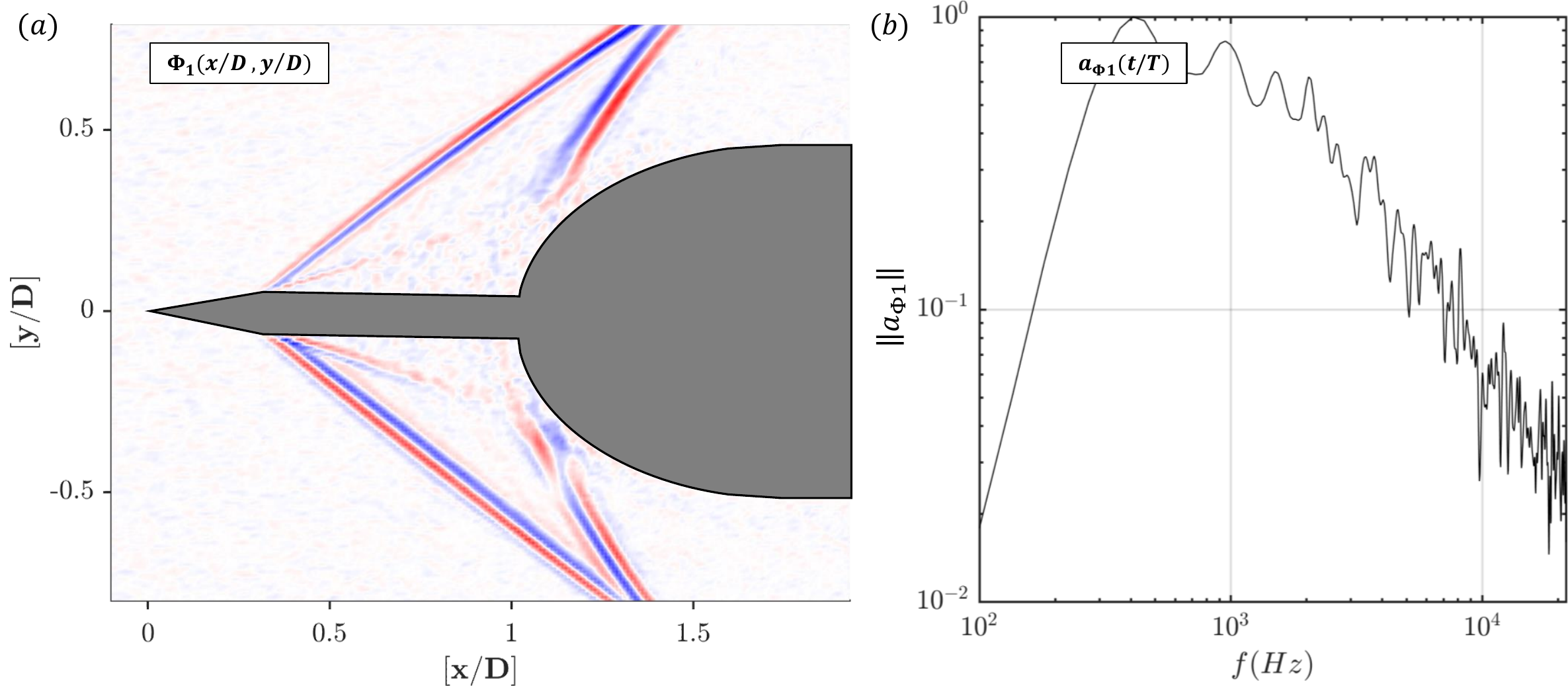}}
		\caption{{(a) Dominant energetic spatial mode ($\Phi_1(x/D,y/D)$) and (b) the normalized spectrum of the temporal mode ($a_{\Phi1}(t/T)$) obtained from the POD analysis of the shadowgraph images for the elliptical spiked body configuration at $M_\infty=2.0$. Flow is from left to right.}}
		\label{fig:26}
	\end{figure} 
	
\begin{figure}
		\centering
		{\includegraphics[width=0.7\columnwidth]{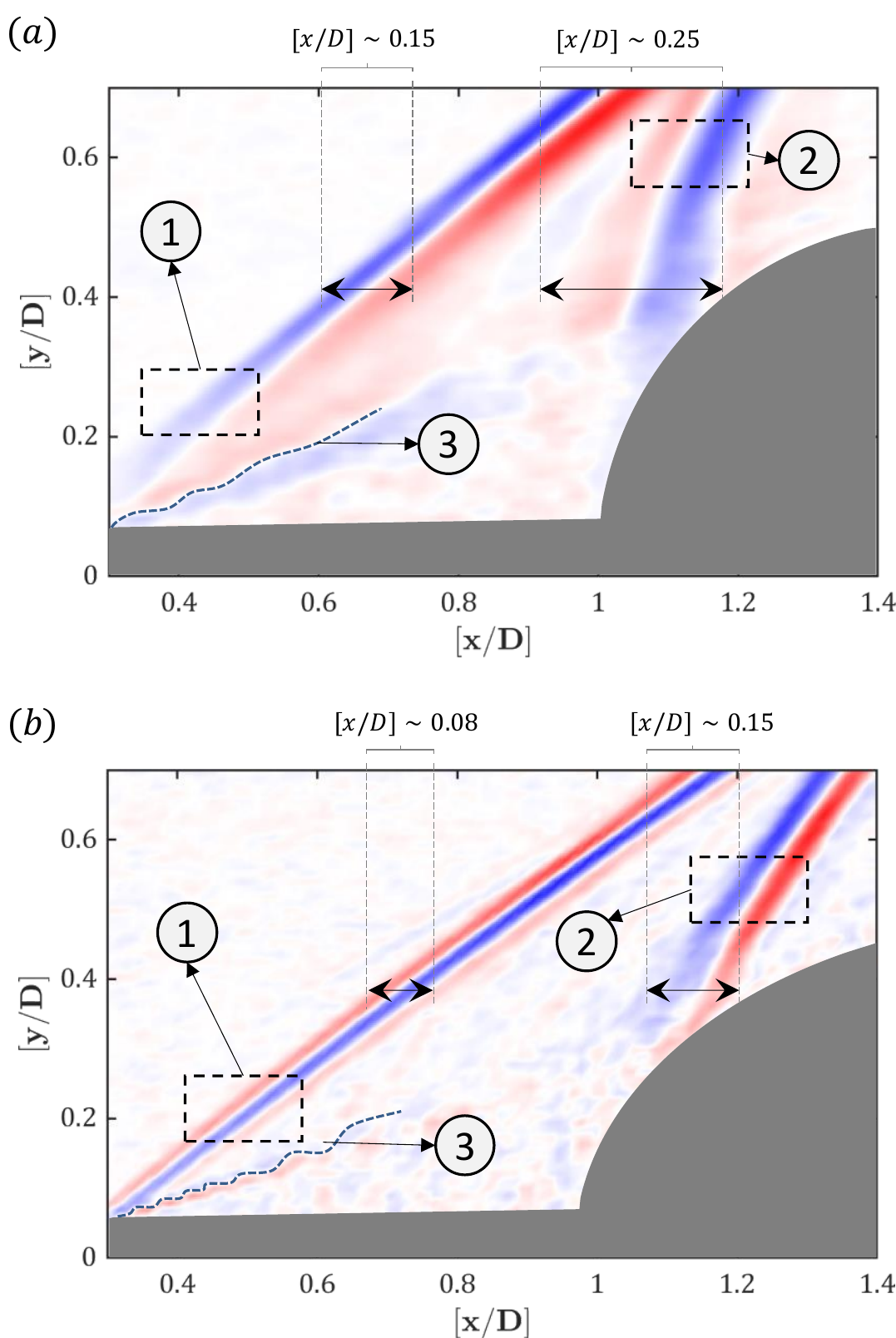}}
		\caption{\hlt{Dominant energetic spatial modes obtained from the POD analysis ($\Phi_1(x/D,y/D)$) for (a) the hemispherical and (b) the elliptical spiked forebody configurations showing the difference in amplitude of out-of-phase shock oscillation at $M_\infty=2.0$. Flow is from left to right. Dominant features: 1. Separation shock (SS), 2. Reattachment shock (RS), 3. Separated free shear layer.}}
		\label{fig:27}
	\end{figure}

	In Figure~\ref{fig:26}, the dominant energetic spatial ($\Phi_{1}(x/D,y/D)$) and temporal modes for the elliptical spiked body configuration are shown. $\Phi_1(x/D,y/D)$ is well correlated to the time-averaged image shown in Figure~\ref{fig:10}c. From Figure~\ref{fig:9}a, one can see that it requires $\sim 100$ modes to represent $\sim$30\% of the energy contents in the flow field, which is comparatively higher than for the hemispherical case ($\sim$70 modes). The requirement for a larger number of modes in the elliptical case is a result of the finer structures in the shear layer in comparison to the hemispherical one. Hence, a broadband spectrum for a wider range of frequencies is seen in the temporal mode of Figure~\ref{fig:26}b. The unsteady pressure spectrum for the elliptical case shown in Figure \ref{fig:11}d (blue line) is also in agreement with the above findings. The power between $\sim 1-2$ kHz is reduced relatively between the hemispherical and the elliptical cases (see blue lines in Figure \ref{fig:11}c-d). The relative drop in power suggests the presence of higher-frequency events in the free shear layer for the elliptical case. The corresponding DMD spectra ($a_\Theta(t/T)$) also show a slight drop in comparison with the hemisphere, confirming to the presence of smaller scales (Figure \ref{fig:9}c). The time-averaged wall-static pressure coefficients (Figure \ref{fig:11}a) and the $x-t$ diagram (Figure \ref{fig:25}) have already shown to be in agreement with the above statements. The difference in the intensity and amplitude of the shock oscillations between the hemispherical and the elliptical spiked body configurations can be more clearly seen in Figure~\ref{fig:27} from the respective $\Phi_1(x/D,y/D)$. It shows the existence of the out-of-phase shocks motion in both cases; however, at a different strength. Such a behavior supports the charging and ejection of fluid mass associated with the recirculation region. The amplitude of shock oscillation decreases by almost 50\% while changing the forebody shape ($\xi$) from the hemisphere to ellipse. 
	
	\subsection{`Aerodisk' flow and unsteadiness control guidelines} \label{case4}
	\hlt{In the hemispherical forebody configuration ($\xi\approx0.92$), the unsteadiness is driven by the separated free shear layer and localized shock oscillations. The cone angle ($\lambda$) of the recirculation region and the out-of-phase shocks motion along with the shocklets affect the size of the large scale structures in the free shear layer. The magnitude of the shock oscillation at the SP (separation point) and RP (reattachment point) are significant. For the elliptical spiked forebody ($\xi\approx0.84$), the mode of unsteadiness is similar to that of the previous case. However, the free shear layer is seen to have finer structures owing to lower $\lambda$ and mild out-of-phase shock motions along with weaker shocklets.}
	
	\begin{figure}
		{\includegraphics[width=\columnwidth]{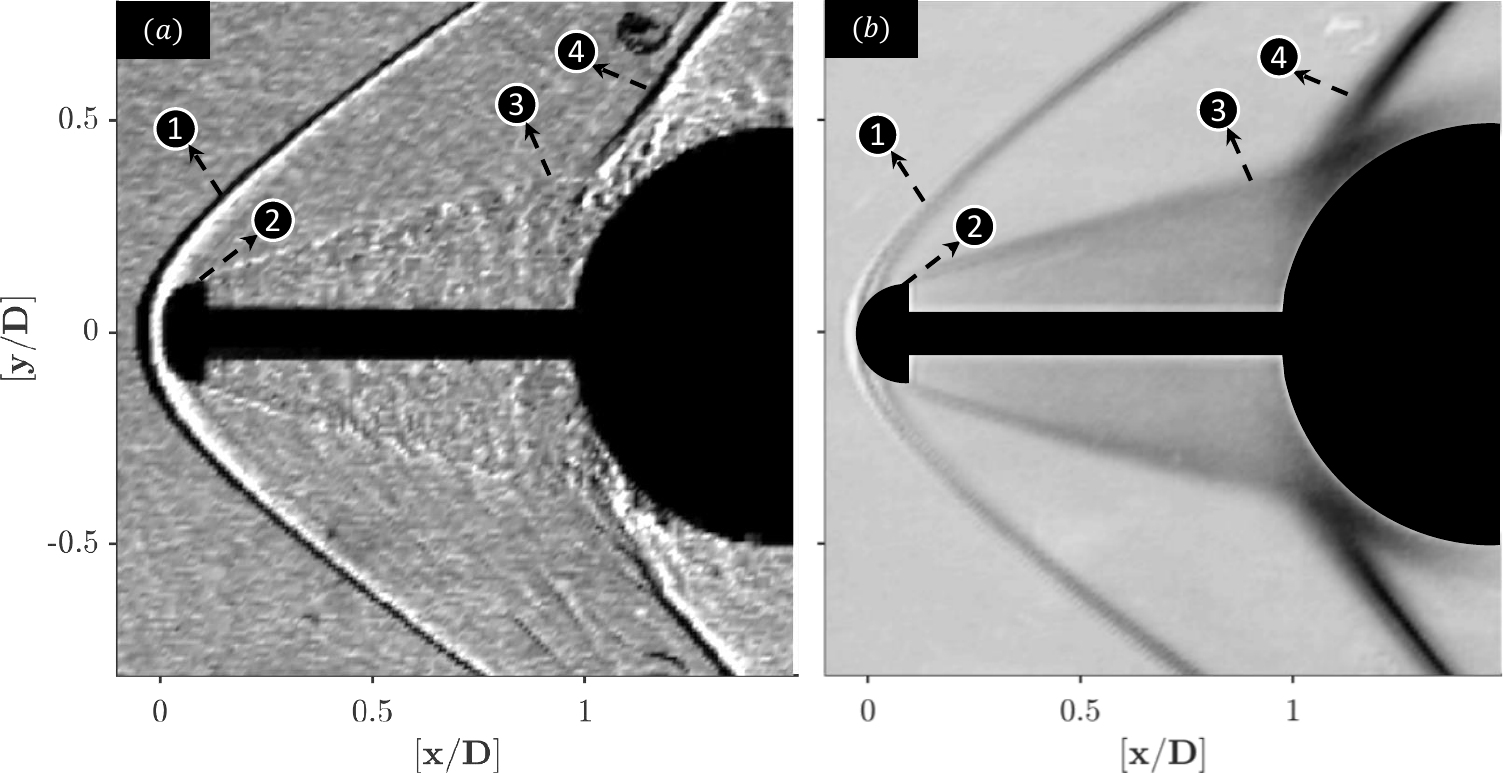}}
		\caption{{(a) Instantaneous shadowgraph image and (b) Operator based time-averaged shadowgraph image ($\left\|\bar{\boldsymbol R}-\boldsymbol R_{rms}\right\|$) of an aerodisk at $M_\infty=2.0$. Flow is from left to right; Dominant flow features: 1. Detached shock ahead of the aerodisk, 2. Separation point (SP, with no separation shock), 3. Recirculation region, 4. Reattachment shock (RS). (Corresponding \href{https://youtu.be/5Ha7kJerZWg}{video}) file is available in the supplementary}}
		\label{fig:28}
	\end{figure} 
	
	\hlt{The present findings can be used to explain the flow past other spiked bodies like the aerodisk configurations which are known to exhibit a reduced level of shock-related unsteadiness. Hence, experiments are conducted on the hemispherical forebody in which the sharp spike tip has been replaced by a hemispherical aerodisk. The diameter of the aerodisk is twice the spike stem $(d)$. A typical instantaneous shadowgraph image and the operator based time-averaged shadowgraph image ($\left\|\bar{\boldsymbol R}-\boldsymbol R_{rms}\right\|$) are shown in Figure \ref{fig:28}. Unlike the previous cases where the unsteady separation shock (SS) is prevalent at the SP, in the present case, the SS is very weak or almost eliminated. There is only a strong bow shock in-front of the aerodisk. Furthermore, from Figure \ref{fig:28}, it seems that the gradient of the gray levels between the free shear layer and its surrounding is much reduced in comparison with that of the sharp-tip spiked-forebodies (Figure \ref{fig:10}b-c). Darker levels suggest stronger activities of the large scale structures along the free shear layer. In the supplementary, a high-speed shadowgraphy \href{https://youtu.be/5Ha7kJerZWg}{video} of the flow past an aerodisk is provided.}
	
	\begin{figure*}
		{\includegraphics[width=0.7\textwidth]{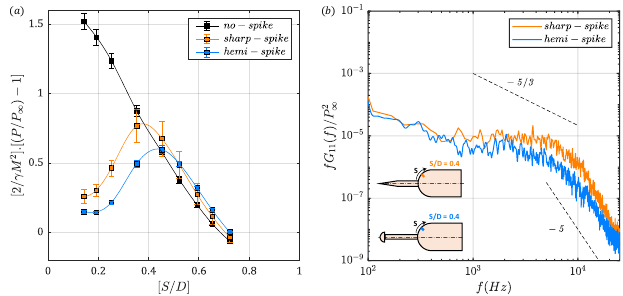}}
		\caption{{(a) Pressure coefficient variation along the forebody curvature length $[S/D]$ for three different hemispherical forebody configurations obtained through time-averaged wall-static pressure measurements; (b) Power spectral density obtained from the unsteady pressure measurement with different spiked tip configurations at $M_\infty=2.0$.}}
		\label{fig:29}
	\end{figure*}
	
	\hlt{Mean wall-static pressure coefficients and unsteady pressure measurements are carried out on the hemispherical spike tip/aerodisk configuration to ascertain the shock dynamics (Figure \ref{fig:29}). Comparison with the no-spike case is helpful to study the flow kinematics around the hemispherical spike tip and the flow conditions in the recirculation region shown in Figure \ref{fig:28}b. Basically, the stand-off shock in-front of the hemispherical forebody and the hemispherical spike-tip is analogous. Hence, the $c_p$ plot provided in Figure \ref{fig:29} for the no-spike case is useful. At about $[S/D]\sim 0.7$, the local pressure reaches the freestream pressure, suggesting that the flow around the shoulder of hemispherical spike tip approaches the sonic velocity ($[M_3/M_1]\lesssim 1$, see Figure \ref{fig:25} for zones identification). The observations are consistent with the experimental studies of a sphere in supersonic flow \citep{Baer1961,Perkins1952}. The approximate flow Mach number just above the separated shear layer is calculated to be around $M_3\sim 1.8$, which is consistent with the $c_p$ shown in Figure \ref{fig:29} for the no-spike case. Computational Mach number values reported in some of the authors previous work \citep{247182:5562234,247182:5562222} are also in agreement with the present findings. At the point of separation (SP) in the hemispherical spike-tip, $M_3$ is larger than the corresponding values of the hemispherical ($M_3=1.63$) and the elliptical ($M_3=1.74$) configurations with a sharp spike-tip. A separated shear layer is formed with $\lambda=17^\circ$ downstream of the SP, due to the geometrical truncation of the aerodisk into a flat rear base (Figure \ref{fig:28}). There is no separation shock (SS) as the flow leaves the aerodisk shoulder tangentially to the separated shear layer. Since, the flow accelerates around the shoulder with $M_3\lesssim M_1$ (supersonic), the separated shear layer cannot oscillate forth and back much. Hence, the disturbances in the recirculation zone have almost negligible influence on the separation point (SP) movement.}
	
	To put everything in perspective, the results are compared against the sharp spike tip case with the same hemispherical forebody. The peak wall static pressure coefficient (Figure \ref{fig:29}a) is smaller and slightly shifted downstream for the aerodisk in comparison with the sharp tip. Such behavior is purely due to the elevation of the separated free shear layer and the consequence of lowering $\lambda$. The spectrum from the unsteady pressure signal obtained at $[S/D]=0.4$ (blue line in Figure \ref{fig:29}b), shows a significant drop in power across almost the entire range of frequencies in comparison to the sharp spike tip case (orange line). The reduction in power, especially at the mid-frequency range, could be attributed to the finer structures and the disappearance of the shocklets in the separated free shear layer.
	
	\begin{figure}
		{\includegraphics[width=0.7\columnwidth]{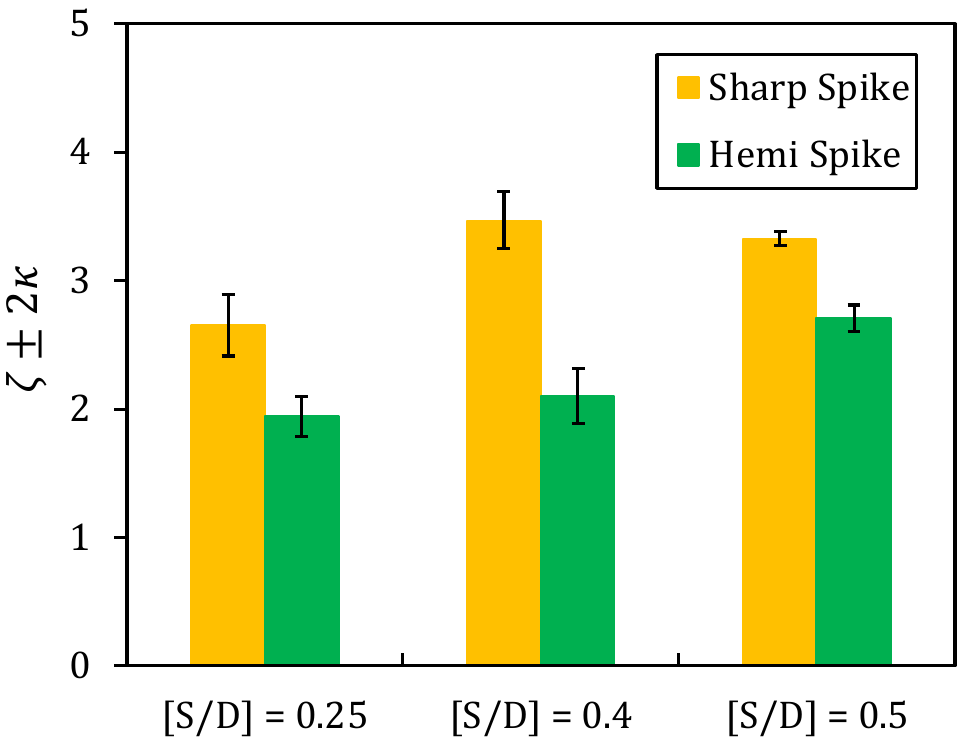}}
		\caption{\hlt{Graph showing the variation of the pressure loading ($\zeta$) and the pressure fluctuation intensity ($\kappa$) parameter at different forebody locations with sharp spike tip, and a hemispherical spike/aerodisk mounted on a hemispherical forebody.}}
		\label{fig:32}
	\end{figure}
	
	\hlt{In Table \ref{tab:3}, values of $c_d$, $\zeta$ and $\kappa$ for the above mentioned cases are given for both the spiked and no-spike models. In the tabulation of Figure \ref{fig:23}, values of $M_C$, $\lambda$, and $\delta'$ are given. Variation of $\zeta$ and $\kappa$ along the forebody surface at three different $[S/D]$ is shown in Figure \ref{fig:32} for the sharp and aerodisk cases. The freestream flow shielding of the hemispherical forebody by the aerodisk by means of a strong bow shock reduces $c_d$ by 10\%. In addition, the aerodisk exhibits a reduction in $\zeta$ by 42\% in comparison to the hemispherical spiked body with a sharp spike tip. Similarly, a reduction in $\kappa$ from 11\% to 10\% is also observed. The larger reduction of $\zeta$ values for the aerodisk at smaller $[S/D]$ suggests minimal shocklets interactions with the spike stem.}
	
	\begin{figure*}
		{\includegraphics[width=0.8\textwidth]{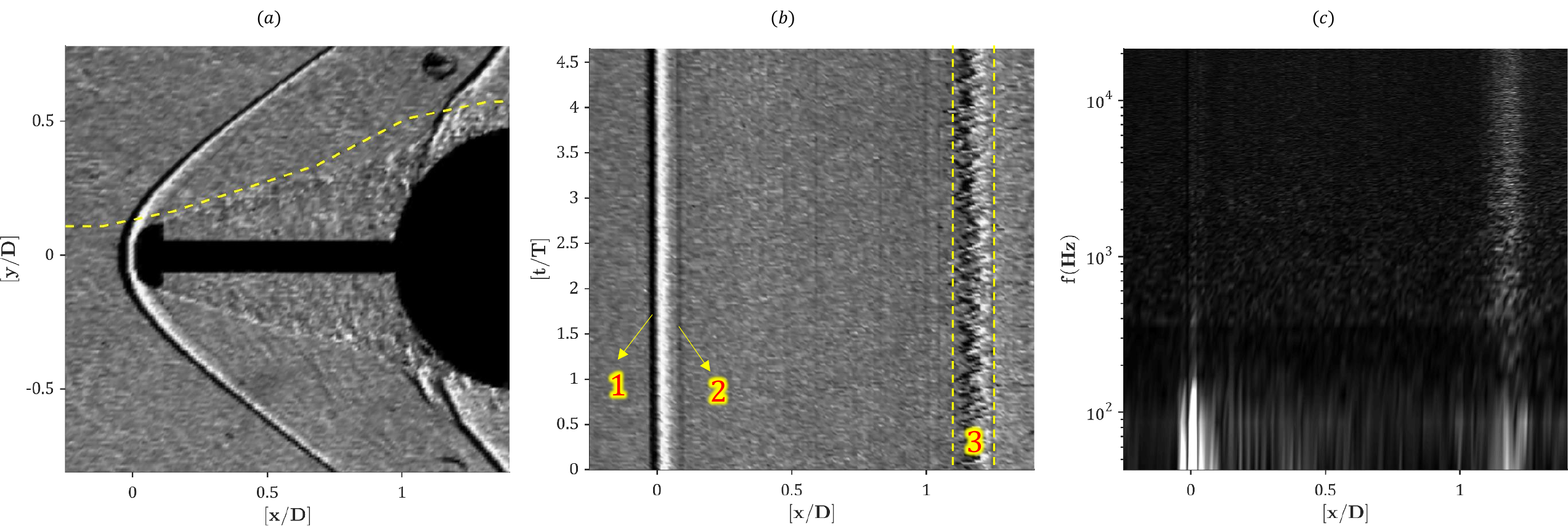}}
		\caption{{Simple sectional image analysis routines performed for the hemispherical spiked body configuration with a hemispherical spike-tip at$M_\infty=2.0$. Flow is from left to right. (a) Yellow-line represents the considered section in a typical instantaneous shadowgraph image; (b) Time evolution (normalized with $T=1$ ms) of the sectional intensity scan revealing the dominant features in the flow: 1. Bow shock in-front of the hemispherical spike-tip, 2. Separated shear layer from the flat base of the hemispherical spike-tip, 3. Oscillating reattachment shock (RS) near the hemispherical forebody; (c) Normalized spectral analysis of the sectional images evolving in time ($x-t$ diagram) revealing the dominant frequency contents in the flow.}}
		\label{fig:30}
	\end{figure*}
	
\begin{figure}
		{\includegraphics[width=\columnwidth]{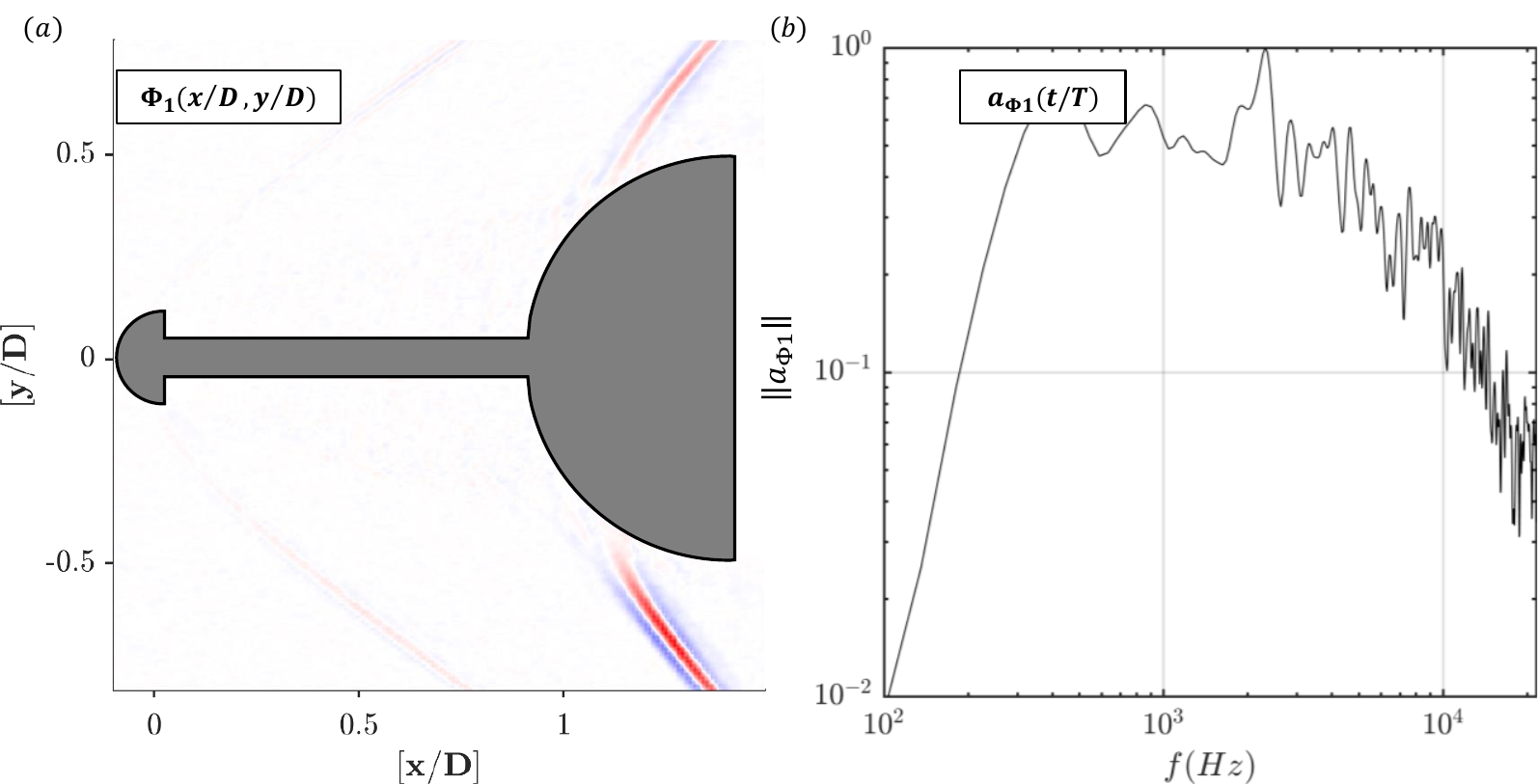}}
		\caption{{(a) Dominant energetic spatial mode ($\Phi_1(x/D,y/D)$), and (b) the normalized spectrum of the temporal modes ($a_{\Phi1}(t/T)$) obtained from the POD analysis of the instantaneous shadowgraph images. The configuration under study is the hemispherical forebody mounted with the hemispherical spike tip at $M_\infty=2.0$. Flow is from left to right.}}
		\label{fig:31}
	\end{figure}

	Shock footprint analysis through the $x-t$ diagram has been carried out to validate the findings (Figure \ref{fig:30}). From the analysis, the bow shock in-front of the hemispherical spike tip and the SP of the free shear layer are observed to be fairly steady. In the reattachment zone, the unsteadiness in the shock motion is contained within $[x/D] \sim 0.12$ which is of the same order as that of the elliptical case ($[x/D] \sim 0.15$) and significantly smaller than the hemispherical case ($[x/D] \sim 0.25$) (see also Figures \ref{fig:18} and \ref{fig:25}). The spectrum of the $x-t$ diagram (Figure \ref{fig:30}c) shows no dominant frequency at the point of separation. Similarly, near the reattachment zone, a low power broadband spectrum is observed. Modal analysis reveals that the dominant energetic spatial mode ($\Phi_{1}(x/D,y/D)$) shows variation only near the RS (Figure \ref{fig:31}a) and there are no distinct events near the SP. Also, unlike in the previous cases (Figure \ref{fig:27}), the alternating color contours near the SP is absent, indicating that there is no out-of-phase shock motion. Finally, in agreement with the $x-t$ diagram, the spectrum from the POD temporal coefficients ($a_{\Phi1}(t/T)$) also reveals to be broadband (Figure \ref{fig:31}b).
	
	\hlt{Thus, it has been demonstrated that the out-of-phase shock motion is almost completely eliminated due to the shape of the spike configuration itself. The point of separation (SP) is shifted vertically above the spike stem and the streamlines leave the hemispherical spike tip/aerodisk tangentially. In addition, $\lambda$ is smaller in comparison with the sharp spike tip case. Hence, for the control of shock-related unsteadiness, the intensity could be reduced or almost completely eliminated by having a streamlined forebody shapes (like elliptical, ogives, or cone-frustum), or by having a thicker spike tip (like aerodisk) or thicker spike stem ($d/D$), which in-turn reduces the separation angle ($\lambda$).}

	\section{Concluding Remarks} \label{conc.}
	\hlt{Experiments are conducted in a supersonic wind tunnel at a freestream Mach number of 2.0 to study the unsteady flow field around three representative spiked forebody configurations. The geometries are selected based upon the forebody geometrical shape factor ($\xi$): (a) Flat-face ($\xi \approx 1$), (b) hemispherical ($\xi \approx 0.92$), and (c) elliptical ($\xi \approx 0.84$) spiked forebody configurations. Time-resolved short-exposure high-speed shadowgraph images and steady and unsteady pressure measurements are carried out to infer qualitative and quantitative details. A shock-foot print analysis using the $x-t$ plots to access the magnitude and time scales of the shock-related unsteadiness. Modal analysis is performed on the time-resolved shadowgraph images to identify the dominant spatiotemporal modes, using Proper Orthogonal Decomposition (POD) and Dynamic Mode Decomposition (DMD) techniques.}  
	
	Following are the major conclusions from the present study:
	
	\begin{enumerate}
		\item \hlt{In case of the flat-face spiked body configuration, for $[l/D]=1$, the well known pulsation mode of unsteadiness \citep{247182:5562227} is observed, producing a significant pressure loading ($\zeta$) and pressure fluctuation intensity ($\kappa$) on the forebody. The `pulsation' cycle is described using the $x-t$ diagram and the modal analysis. The spatiotemporal modes from the shadowgraph images are captured for the first time, and are validated against previous findings.}
		
		\item In the case of the hemispherical spiked body configuration, the shock-related dynamics are dictated by the flow dynamics of separated free shear layer and the recirculation region. The charging and ejection of fluid mass from the recirculation region, coupled with the out-of-phase shocks motion, is the driving flow physics. Unlike the flat-case, here the power spectral distribution is broad. 
		
		\item The shock-related unsteadiness in the elliptical case is the lowest and the flow physics is similar to that of the hemispherical case. The lower cone angle ($\lambda$) of the recirculation region plays an essential role in reducing unsteadiness.
		
		\item Flow physics: As $\lambda$ increases, the shocklets along the free shear layer become stronger. Stronger shocklets appear to undergo multiple reflections and refractions inside the recirculation region. The resulted disturbances are deflected upstream and perturb the separation shock (SS). The perturbed SS oscillates and further disrupts the separated shear layer. The disruptions get amplified by the KH instabilities in the shear layer, and large scale structures are formed. Further downstream, they grow in size and impinge at the reattachment point (RP). The large scale structures contribute to the charging and ejection of fluid mass inside the recirculation region. Consequently, during ejection, the reattachment shock (RS) oscillates and enables the out-of-phase motion of SS with a magnitude proportional to the size of the large scale structures.
		
		\item \hlt{Streamlined forebody shape (lower $\xi$), elimination of out-of-phase shock motion, and lower $\lambda$ are found to be the desired characteristics to reduce shock-related unsteadiness.}
		
		\item \hlt{The well-known hemispherical spike tip/aerodisk configuration is shown to have a lower $\lambda$ and almost an eliminated out-of-phase shock motion, thereby reducing significantly the shock-related unsteadiness }
		
		\item \hlt{Unsteadiness control guidelines include having a streamlined forebody shapes (like elliptical, ogives, or cone-frustum), or by having a thicker spike tip (like aerodisk) or thicker spike stem ($d/D$), which in-turn reduces the separation angle ($\lambda$).}
	\end{enumerate}
	
	\subsection*{Acknowledgments}
	The first author thanks the Technion research funding during his doctoral studies. The second author gratefully acknowledge the Post-Doctoral Fellowship offered in parts by the Fine Trust along with Technion. The authors would like to thank the help of Mark Koifman, Michael Dunaevsky, Oleg Kan, Yafim Shulman, Nadav Shefer, and David Naftali in conducting the experiments, effectively.
	
	The authors report no conflict of interest.
	
	\begin{appendix}
		\counterwithin{figure}{section}
		\section{Data analysis using modal decomposition}\label{data_an}
		Firstly, the obtained shadowgraph images from the procedure mentioned in $\S$ \ref{shadow} are subjected to a rigorous image pre-processing routines using Matlab programs. During the acquisition, image intensity saturation is avoided to monitor the fluctuations at all spatial points.  Due to the non-uniformity and defects present in either the optical windows or the mirrors, the obtained images may possess spatial intensity variations. Besides, the density variations along the light path inside the laboratory environment cause spatiotemporal light intensity fluctuations. These anomalies need to be sorted out. Else the analysis picks up these features as the dominant ones. Many investigators (e.g. \cite{kutz,prothlin}) have followed some standard image pre-processing routines. In the present study, the time series subtraction filter and the image intensity normalization filter used by \unskip~\cite{247182:5751914} in the acetone planar laser-induced fluorescence (acetone-PLIF) studies are used to prepare the shadowgraph images. Besides, the images are also corrected for perspective distortions and vignetting effects arising due to the usage of the telephoto type lens system. Later, the corrected images are calibrated based upon the physical scale of the model geometry.
		
		In the present paper, data-driven analysis techniques like Proper Orthogonal Decomposition (POD) \citep{247182:10033594,247182:10033592,247182:5609084,247182:5609089} and Dynamic Mode Decomposition (DMD) \citep{247182:5609095,247182:5741225}, are used to identify the dominant energetic and dynamic contents in the flow field. The authors have used the classic decomposition tool called `method of snapshots' \citep{247182:5609091,247182:5609084} to identify the dominant modes from the shadowgraph images. Traditionally these methods are used on vector fields \citep{247182:5609089} obtained from PIV analysis. However, it has also been utilized in the scalar fields \citep{247182:5609086,247182:5609095}. The methodology employed in the present study is adopted and expanded from the procedure described in the work of \unskip~\cite{247182:5609093}, and \unskip~\cite{247182:5609094,247182:5609092}. For simplicity a brief mathematical routine is described here.
		
		\begin{align}
			\label{pod1}
			\boldsymbol{R}&=\{\boldsymbol r_1,\boldsymbol r_2,...\boldsymbol r_n\},\\
			\label{pod2}
			\boldsymbol{\overline{R}} &=\frac{1}{n}\sum_{1}^{n} \boldsymbol r,\\
			\label{pod3}
			\boldsymbol{R}' &=\boldsymbol{R}-\boldsymbol{\overline{R}} =\{\boldsymbol r'_1,\boldsymbol r'_2,...\boldsymbol r'_n\}.
		\end{align}
		
		Each snapshot of size $i \times j$ (in pixels) carrying the light intensity fluctuations represents the line of sight integrated variations in the double derivative of the flow field density ($\sim \boldsymbol \nabla^2 \rho$), directly. The obtained two-dimensional snapshot is represented as a column vector $\left(\boldsymbol r\right)$ of size $k=i \times j$. A large matrix $(\boldsymbol{R})$ based on the total number of snapshots ($n$) is thus formed by converting and stacking the temporally evolving images in succession. The final size of the snapshot matrix is $k \times n$. After utilizing the image pre-processing routines, the instantaneous images $\left(\boldsymbol R\right)$ are used to create a time-averaged image $\left(\overline{\boldsymbol R}\right)$. A sequence of images carrying the fluctuation field alone $\left(\boldsymbol R'\right)$ are then computed. The steps are briefly described in equation \ref{pod1}-\ref{pod3}. The value of $n$ in the current studies is 1000 as no significant difference was found by increasing $n$ from 1000 to 10000. The snapshots are captured at an image resolution of 256 $\times$ 160 pixels at 43000 Hz with 2 $\mu$s light exposure (see Figure \ref{fig:6}). Thus, the spatial and spectral resolution in the observed dynamic and energetic spatiotemporal modes is 0.5 mm/pixel and 43 Hz, respectively.
		
		\begin{align}
			\label{pod4}
			\boldsymbol{R}'{_1} &=\{\boldsymbol r'_1,\boldsymbol r'_2,...\boldsymbol r'_{n-1}\},\\
			\label{pod5}
			\boldsymbol{R}'{_2} &=\{\boldsymbol r'_2,\boldsymbol r'_3,...\boldsymbol r'_n\},\\
			\label{pod6}
			\boldsymbol R'_2 &= \boldsymbol A \boldsymbol R'_1.
		\end{align}
		
		Two time-lagged snapshot matrices ($\boldsymbol R'_1, \boldsymbol R'_2$) are constructed by only considering snapshots between $1$ to $n-1$ and $2$ to $n$ in order to study the time dynamics (see equation \ref{pod4} and \ref{pod5}). The evolution of a time-dependent system of variables like $\boldsymbol R$ in terms of a linear propagator $\boldsymbol{A}$ is represented through time-lagged matrices ($\boldsymbol R'_1, \boldsymbol R'_2$) as stated in equation \ref{pod6}. Solving for $\boldsymbol{A}$ is computationally expensive and hence, it has been approximated using a companion matrix $\widetilde{\boldsymbol B}$. Later, one of the time-lagged matrix ($\boldsymbol{R'_1}$) is rewritten through a singular value decomposition (SVD) as stated in equation \ref{pod7} and \ref{pod8} \cite{247182:5741225}. 
		
		\begin{align}
			\label{pod7}
			\boldsymbol R'_2 &\approx \widetilde{\boldsymbol B} \; \boldsymbol R'_1,\\
			\label{pod8}
			\boldsymbol R'_2 &\approx {\boldsymbol U} {\boldsymbol S} {\boldsymbol V^H} \widetilde{\boldsymbol B}.
		\end{align}
		
		In the above stated equations, equation \ref{pod8} carries the energetic spatiotemporal modes obtained from the singular value decomposition of the time-lagged matrix $R'_2$ \citep{247182:5741225}. Hence, the SVD step can be considered as a POD operation, since the decomposition results in the formation of $\boldsymbol U$ containing the spatial information $\boldsymbol \Phi_n$, $\boldsymbol V$ having the temporal information $\boldsymbol a_{\Phi n}(t)$, and $\boldsymbol S$ has the energy contents $\boldsymbol \lambda$ (eigenvalues). It has to be noted that the singular values ($\boldsymbol S$) and the eigenvalues ($\boldsymbol \lambda $) are related by $\boldsymbol S^{2}=\boldsymbol \lambda $ and hence the percentage of energy contents in each mode has to be calculated accordingly \citep{taira}. While using shadowgraph images, the word energy is, in general, a misnomer, as the eigenvalues carry only the square of light intensity fluctuations, unlike the more conventional use of the square of velocity fluctuations obtained from velocimetry data. However, for the consistent usage with the existing literature, the terminology is \hlt{kept the same} in this manuscript.
		
		\begin{align}
			\label{pod9}
			\widetilde{\boldsymbol B} &\equiv {\boldsymbol U^H} \boldsymbol R'_2 {\boldsymbol V} {\boldsymbol S^{-1}},\\
			\label{pod10}
			\widetilde{\boldsymbol B} \boldsymbol \Lambda &= \boldsymbol \Delta \boldsymbol \Lambda.
		\end{align}
		
		After performing the decomposition, the approximated matrix ($\widetilde{\boldsymbol B}$) is evaluated as stated in equation \ref{pod9} and subjected to an eigenvalue problem formulation like in the case of a POD analysis \citep{247182:5609089} as given by equation \ref{pod10}. In equation \ref{pod10}, $\boldsymbol \Delta$ represents the eigenvector and $\boldsymbol \Lambda$ contains the eigenvalues. The DMD spatial modes ($\boldsymbol \Theta$) are then computed as shown in equation \ref{pod11}. The DMD spectral contents ($\boldsymbol{f_{\boldsymbol{\Theta}}}$) are accessed by taking the imaginary part of the products of $f_s/2\pi$ and $\log(\boldsymbol{\Lambda})$, where $f_s$ is the sampling rate at the time of image acquisition (see equation \ref{pod12}). The DMD amplitudes ($\alpha_{\boldsymbol{\Theta}}$) are obtained by taking the pseudo inverse of any one of the DMD modes ($\boldsymbol \Theta$) and projecting it on to the corresponding fluctuating snapshot ($\boldsymbol r'$) from the respective time-lagged column matrix ($\boldsymbol R'_2$) as stated in equation \ref{pod13}. In equation \ref{pod13}, the first DMD spatial mode ($\boldsymbol\theta_1$) and the first fluctuating snapshot ($\boldsymbol r'_2$) are used to compute the amplitude ($\alpha_{\boldsymbol{\Theta}}$). However, such a routine for a larger matrix will be computationally expensive. For a more efficient method to compute the DMD amplitudes, the readers are referred to the paper of \unskip~\cite{247182:5741684}.
		
		\begin{align}
			\label{pod11}
			{\boldsymbol \Theta} &= \{\boldsymbol \theta_1,\boldsymbol \theta_2,...\boldsymbol \theta_{n-1}\} 
			= \boldsymbol R'_2 \boldsymbol V \boldsymbol S^{-1} \boldsymbol \Delta, \\
			\label{pod12}
			\boldsymbol{f_{\boldsymbol{\Theta}}} &= \operatorname{\mathfrak{I}}\left(\frac{f_s \log{(\boldsymbol \Lambda)}}{2\pi}\right),\\
			\label{pod13}
			\alpha_{\boldsymbol{\Theta}} &= \boldsymbol\theta_1^+ \boldsymbol r'_1.
		\end{align}
		
\begin{figure*}
			\centering{\includegraphics[width=0.9\textwidth]{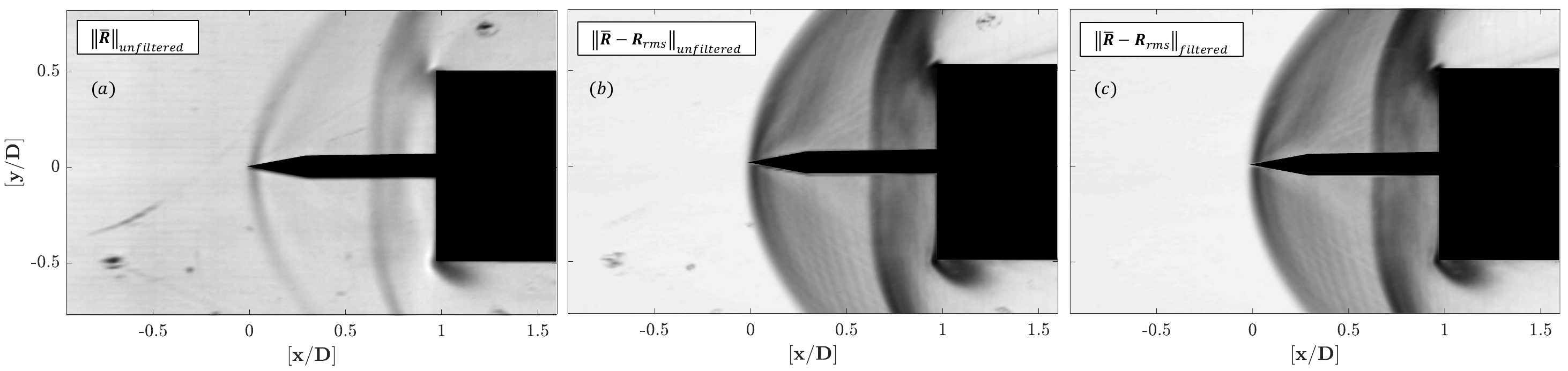}}
			\caption{{Time-averaged shadowgraph imaging of a flat-face spiked body configuration at $M_\infty=2.0$: (a) $\left\|\bar{\boldsymbol R}\right\|$ from the unfiltered image sets, (b) $\left\|\bar{\boldsymbol R}-\boldsymbol R_{rms}\right\|$ from the unfiltered image sets, and (c) $\left\|\bar{\boldsymbol R}-\boldsymbol R_{rms}\right\|$ from the filtered image sets using the DMD routines prescribed by \cite{kutz}. Flow is from left to right.}}
			\label{fig:8}
		\end{figure*}

		With the DMD components $\boldsymbol{\Theta}$, $\boldsymbol{f_{\boldsymbol{\Theta}}}$ and $\alpha_{\boldsymbol{\Theta}}$, unwanted image noise or irrelevant modes (spatiotemporal artifacts) from the snapshots are identified. Higher values of $\alpha_{\boldsymbol{\Theta}}$ for lower $\boldsymbol{f_{\boldsymbol{\Theta}}}$ in the DMD spectrum generally corresponds to parasite reflections from the scratches on the viewing window, and they are removed after visualizing the corresponding DMD components. The time-lagged matrix is approximated and reconstructed using the remaining DMD components for the considered time period. The time evolution term ($\boldsymbol t$) and the eigenvalue term ($\boldsymbol\Lambda$) are used along with the other DMD components to reconstruct the approximated time-lagged matrix ($\widetilde{\boldsymbol R}'_2$) as described in equation \ref{pod14} and \ref{pod15}. More details about the DMD based image noise removal is available in the book of \cite{kutz}. Similar image noise filtering routines to those described in the above mentioned book, are carried out in the present analysis. The difference between the unfiltered and filtered operator-based time-averaged image for the flat-face spiked configuration is demonstrated in Figure~\ref{fig:8}. 
		
		\begin{align}
			\label{pod14}
			\boldsymbol{t} &= \left\{\frac{1}{f_s},\frac{2}{f_s},...\frac{n-1}{f_s} \right\},\\
			\label{pod15}
			\widetilde{\boldsymbol R}'_2 &\equiv {\boldsymbol \Theta}  \left[\exp\left(f_s\log{(\boldsymbol{\Lambda})}\boldsymbol{t}\right)\alpha_{\boldsymbol{\Theta}}\right].
		\end{align}
		
		In the unfiltered normalized time-averaged image (Figure \ref{fig:8}a), scratches in the viewing window, readout lines from the camera sensor (light horizontal lines), and the dynamic flow events around the body are visible with similar light intensity. Hence, the actual flow events around the body are not distinguished. However, while performing a normalized operator-based time-averaging (Figure \ref{fig:8}b), where the normalized difference between $\boldsymbol{\bar R}$ and $\boldsymbol R_{rms}$ is considered, dynamic flow events around the flat-face spiked configuration is distinctly visible. The window scratches that cause light intensity fluctuations are still present in the imaging frame, irrespective of the operator-based time-averaging routines. After performing DMD based image noise filtering (see Figure \ref{fig:8}c), it is evident that only the dominant flow events around the body of interest are existing and all the other noises are filtered out, including the window defects. In the present studies, all the images are subjected to a thorough DMD based image noise filtering routines, before they are considered for any analysis.
		
		\begin{figure}
			\centering{\includegraphics[width=\columnwidth]{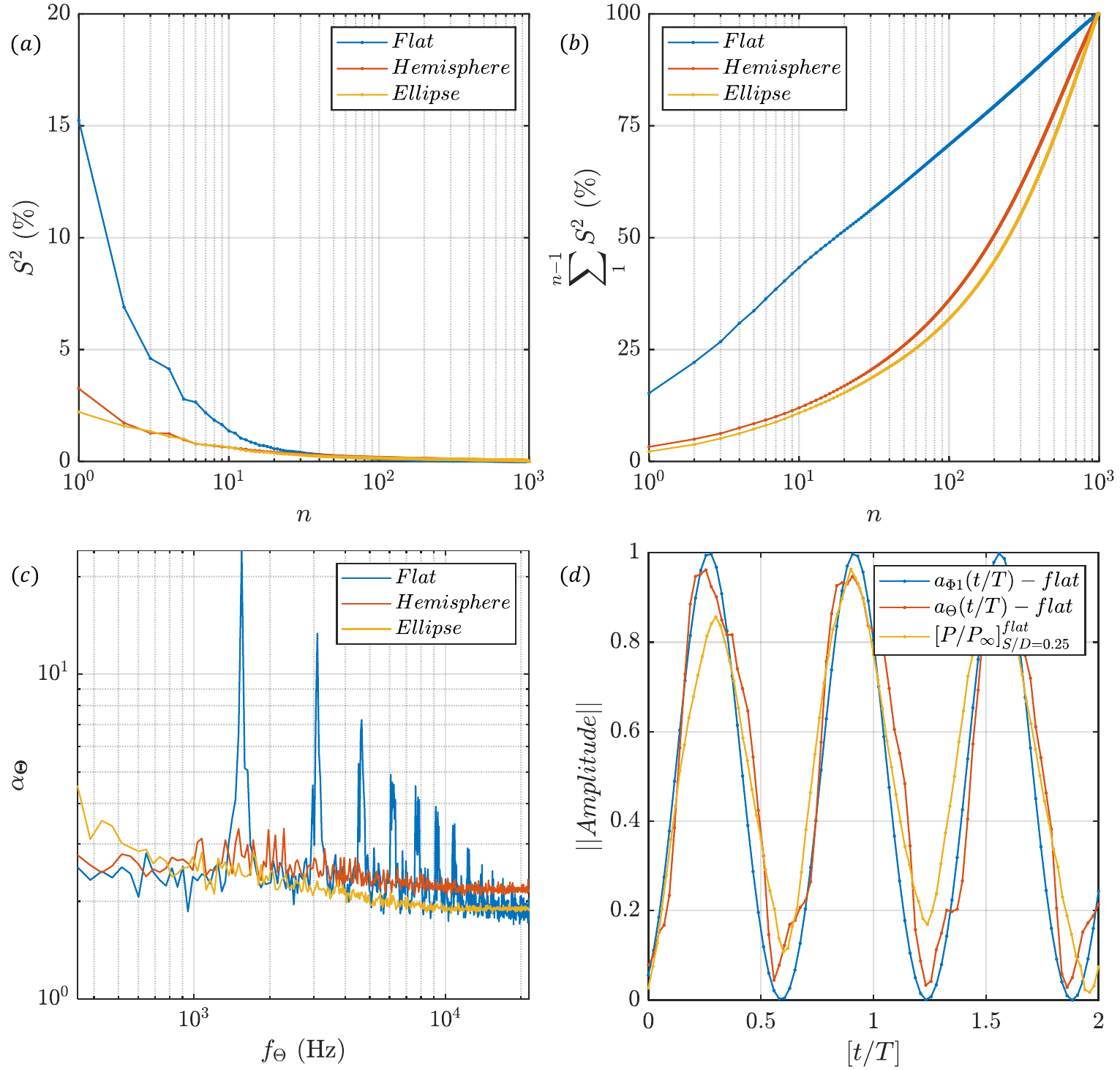}}
			\caption{{(a) The energy contents (in \%) in each mode observed for the different spiked body configurations obtained from the POD analysis; (b) The associated cumulative POD energy contents for the different spiked body configurations; (c) Spectra for the different spiked body configurations obtained from the DMD analysis; (d) Comparison of the POD time coefficients for the first mode $(a_{\Phi1}(t/T))$, and the DMD temporal signal $(a_{\Theta}(t/T))$, with a static pressure signal measured at a particular location $([P/P_{\infty}]_{S/D=0.25})$ for the flat-face spiked forebody configuration.}}
			\label{fig:9}
		\end{figure}
		
		Few of the POD and DMD parameters are compared both qualitatively and quantitatively with the experiments to validate the data analysis techniques. The energy contained in each of the modes ($\boldsymbol{S}^2$) for all the spiked body configurations obtained from the POD analysis is plotted in Figure~\ref{fig:9}a using equation \ref{pod8}. The first ten modes ($n=10$) are found to be dominant, and they influence the overall flow behavior. The rest of the modes contain only a little fraction of the energy in each of the three cases. The first few modes of the flat-face spiked body configuration contain a larger fraction of energy in comparison with the other configurations. In Figure \ref{fig:9}b, the cumulative energies $\left(\sum_{1}^{n-1} \boldsymbol{S}^2\right)$ associated with each of the modes ($n$) are shown. 40\% of the cumulative energy associated with the flat-face spiked body is contained in the first ten modes ($n=10$), whereas the other cases require at least $n=100$ (hemispherical spiked body) and $n=200$ (elliptical spiked body) to represent the same amount of energy.
		
		The drastic difference in the cumulative energy between the flat-face and the other cases is obvious while seeing Figure \ref{fig:6}, and the corresponding videos given in the supplementary. The close resemblance between the flow features in Figure \ref{fig:6}b and Figure \ref{fig:6}c explains the similar order of $n$ to represent 40\% of the cumulative energy in the hemispherical and elliptical cases, respectively. The DMD components $\alpha_{\boldsymbol{\Theta}}$, and $\boldsymbol f_{\boldsymbol\Theta}$ are plotted for all the cases in Figure \ref{fig:9}c. The flat-face case exhibits periodic fluctuations, whereas the other cases are results of broadband spectra. Comparing the wall-static pressure signals at $[S/D]=0.25$ for the flat-face case $\left([P/P_{\infty}]\right)$ with the corresponding dominant POD signal ($a_{\Phi 1}(t/T)$) and the decomposed DMD signal ($a_{\Theta}(t/T)$) shows good agreement as seen in Figure \ref{fig:9}d, thus validating the present analysis. More observations and discussions on the flow field analysis using the POD and DMD techniques are given elaborately in $\S$ \ref{rd3}.
		
		\section{Flow past the flat-face spiked configuration}\label{flow_flat-face}
		
		In Figure~\ref{fig:13}a, the strong shock wave starts to accelerate downstream from the leading edge of the spike ($S1_{in}$) towards the flat-face body till $[x/D]\approx 0.6$. The velocity ($u/U_\infty$) and location ($x/D$) of $S1_{in}$ are easily extracted from the $x-t$ diagram (Figure \ref{fig:15}). The acceleration of $S1_{in}$ is seen in the initial frames from t to t+2$\triangle$t (Figure \ref{fig:13}a) and also in Figure \ref{fig:15}c, thus initiating the phase of `collapse'. When $S1_{in}$ passes through the spike stem, the supersonic free stream flow ($U_{\infty}$) creates a weak shock (WS) from the sharp spike tip (t+$\triangle$t). The recirculating flow from the previous cycle (seen as turbulent structures downstream of $S1_{in}$) convects along the downstream at a supersonic velocity just like $S1_{in}$. As $S1_{in}$ accelerates towards the forebody, the convecting recirculation region (CRR) gets compressed in front of the forebody (from t to  t+3$\triangle$t). From Figure \ref{fig:15}c, it could be seen that the velocity reaches to $[u/U_{\infty}]\approx 1$ at $[x/D]\approx 0.6$. Such higher velocities result in the formation of a strong detached shock ($S2_{out}$) in front of the flat-face forebody (see frames from t+4$\triangle $t to t+6$\triangle $t). The produced detached shock wave ($S2_{out}$) gains strength and moves upstream gradually till $[x/D]\approx 0.6$ at $[u/U_{\infty}]\approx -0.1$ (Figure \ref{fig:15}c) due to the rapid compression of CRR by $S1_{in}$ against $S2_{out}$. As $S1_{in}$ and $S2_{out}$ move against each other, a weak compression front (WF) is formed in-between around $[x/D]\approx 0.6$ (t+6$\triangle $t and t+7$\triangle $t). Expansion fan (EF) is seen near the shoulder of the sharp spike tip after $S1_{in}$ passes through it (t+7$\triangle$t). The flow in front of $S2_{out}$ separates according to $\beta-\theta-M$ criterion in deflecting the flow downstream of $S2_{out}$. The deflected flow passes around the forebody and enters into the freestream. The flow separation results in the formation of lambda shock foot (SF, t+8$\triangle$t). Later, $S2_{out}$ collides with the incoming $S1_{in}$ (from t+7$\triangle $t to t+9$\triangle $t) and thus, ending the phase of `collapse'.
		
		The phase of `inflation' begins after the head-on collision of the two shocks $S1_{in}$ and $S2_{out}$ (t+10$\triangle$t). The head-on collision leads to the growth of SF with a triple point (TP), a front shock foot (FSF), and a rear shock foot (RSF) (from t+10$\triangle$t to t+11$\triangle$t). The shock wave boundary layer interaction (SWBLI) leads to the inflation of FSF, and the flow later evolves in the same way as observed in shock tubes \citep{247182:8346723}. After $S1_{in}$ moves through the lambda shock foot system and hits $S2_{out}$, it gets refracted. The refracted shock (RS) hits the flat-face forebody (t+12$\triangle $t) causing it to produce harmonics in the unsteady pressure spectra (see Figure \ref{fig:11}b). The FSF is considered to be the dominant shock and continues to move upstream towards the spike tip as $S1_{out}$ with decreasing velocity (Figure \ref{fig:15}c). A shear layer interface (SLI) exists between the separated flow region and FSF (t+14$\triangle $t). The vortices shedding from the TP (VSTP) interacts with RS and starts to feed the flow below SLI (from t+13$\triangle $t to t+19$\triangle $t) and inflates FSF or $S1_{out}$. However, RSF hangs between TP and SLI (t+18$\triangle$t). The inflating recirculating region (IRR) is filled with turbulent structures in t+16$\triangle$t. The clockwise VSTP forms a toroidal vortex (TV), and the eye of TV is seen in the toroidal vortex ring (TVR) as dark spots (t+17$\triangle$t). During this process, the formed TP starts to move in the vertical direction (from t+15$\triangle$t to t+19$\triangle$t). 
		
		Due to the line of sight integration in the shadowgraph imaging, the locus of TP and RSF is seen as a vertical line (t+18$\triangle$t); however, it is just an axisymmetric ring. Later, LTP and RSF are considered to be evolving as $S2_{in}$. The phase of inflation comes to an end when FSF or $S1_{out}$ is distinguishable from lambda shock foot and turns to a bow shock at the leading edge of sharp spike tip (t+19$\triangle$t).
		
		During the phase of `withhold,' the curvature of $S1_{out}$ changes further, and it nearly becomes a normal shock (from t+20$\triangle $t to t+29$\triangle $t). The flow fed below the SLI keeps inflating the recirculating region, and it holds $S1_{out}$ at the leading edge of the sharp spike tip. Owing to the change in curvature of $S1_{out}$, $S2_{in}$ is pushed towards the forebody (frame t+22$\triangle$t to t+29$\triangle$t). The charging of flow by VSTP continues until the height of SLI exceeds the value of $D/2$ (t+24$\triangle$t) from the axis. Once $S2_{in}$ reaches near the forebody and SLI crosses the forebody height ($D/2$) with a zero slope, the trapped IRR is pushed back to the freestream. The recirculating region starts to convect (CRR) along the freestream direction, and $S1_{out}$ can no longer be held at the sharp spike tip (Figure \ref{fig:15}c). The  \hlt{upheld} $S1_{out}$ at the spike tip begins to collapse and starts accelerating towards the forebody as $S1_{in}$. Thus, the pulsation cycle repeats. In this phase, the velocity of $S1_{in}$ and $S2_{in}$ is of a similar trend (Figure \ref{fig:15}c). The velocity of $S2_{in}$, however, reaches to higher values ($[u/U_{\infty}]\approx 1.2$ at $[x/D]\approx$ 1.4) more than that of the freestream for a shorter duration since it has just crossed the expansion corner.
		
	\end{appendix}
	
	\bibliography{PRF}
	\bibliographystyle{unsrtnat}
	
\end{document}